\documentclass[12pt]{article}%
\usepackage{amsfonts}
\usepackage{amsmath}
\usepackage{latexsym}
\usepackage{amssymb}
\usepackage{graphicx}%
\providecommand{\U}[1]{\protect\rule{.1in}{.1in}}
\begin{document}

\begin{center}
{\Large A self-confined Fermi-gas model for nuclear collective motion
}\ \bigskip

\ V.P. Aleshin

Institute for Nuclear Research, Kiev, 03680, Ukraine

\bigskip

Abstract
\end{center}

The well-known phenomenological dynamic equation for the nuclear shape
parameter was derived from first principles with the aim of obtaining the
microscopic or many-body expressions for inertia $B$, deformation force $f$,
and friction coefficient $\gamma$ of collective motion in hot nuclei. Nuclear
evolution, viewed as a sequence of quasiequilibrium stages, is described with
the aid of the density matrix $\rho_{\mathrm{q}}$ constrained to given
expectation values of nuclear Hamiltonian $\mathcal{H}$, number operator
$\mathcal{N}$, and operators $Q$, $P$, and $\mathcal{M}$ of coordinate,
momentum, and inertia of collective motion. Having chosen an explicit
expression for $Q$ in terms of the nucleon field operators $\psi_{\mathbf{x}}%
$, $\psi_{\mathbf{x}}^{\ast}$, we construct the corresponding expressions for
$\rho_{\mathrm{q}}$, $P$, and $\mathcal{M}$, using a certain canonical
transformation of $\psi_{\mathbf{x}}$, $\psi_{\mathbf{x}}^{\ast}$, the
equation of continuity and an assumption that the collective variables
$p_{n,t}=\mathrm{tr}\left(  P_{n,t}\rho_{\mathrm{q}}\right)  $ where $P_{n,t}$
are Heisenberg representations of $P_{1}=Q$, $P_{2}=P$, $P_{3}=\mathcal{M}$,
change with time much slower than the number density and the momentum density,
from which $p_{n,t}$ are built. The same assumption was used to get the
closed-form solutions of the dynamic equations for $p_{n,t}$, from which we
extract the desired many-body expressions for $B$, $f$, and $\gamma$. After
adaptation of those general expressions to a self-confined Fermi gas with the
phenomenological effective force, they are used for critical analysis of the
previous microscopic models of those quantities and for elucidating the
distinctive features of dissipative collective motion in atomic nuclei.

\ PACS numbers: 25.70.Lm, 24.10.Cn, 25.70.Jj

\section{Introduction}

Collective or 'macroscopic' motion of atomic nuclei consists in large-scale
deformations of nuclear shape, characterized by the parameter $q=q\left(
T\right)  $, which satisfies the equation of motion
\begin{equation}
B\ddot{q}+\frac{1}{2}\dot{q}{}^{2}\partial_{q}B=f-\gamma\dot{q}, \label{deI}%
\end{equation}
with $\dot{q}=\partial q/\partial T$, $\ddot{q}=\partial^{2}q/\partial T^{2}$,
in which inertia $B$, deformation force $f$, and friction $\gamma$ are certain
functions of $q$. Suggested in \cite{BW,Kr,HW} for neutron-induced fission,
Eq. (\ref{deI}), after coupling to particle evaporation from strongly deformed
nuclei \cite{Aleshin96} and completing by a random force \cite{FG,RKNA}, has
become a promising tool for an in-depth analysis of experimental data on
fusion-fission, fusion-evaporation and quasifission processes following
nucleus-nucleus collisions at the energies per nucleon sizably below the
Fermi-energy \cite{Gavron87,Hinde92,BFUSAA,RTH}.

The microscopic content of $B$, $f$, $\gamma$, and $q$ is obscure, because Eq.
(\ref{deI}) has never been derived, as a whole, from quantum-statistical
theory of self-bound nonequilibrium Fermi systems. This work suggests such a
derivation, under hypothesis that Eq. (\ref{deI}) is a consequence of the
dynamic equations for 'macroscopic', 'gross' or collective variables of the system.

The first step towards microscopic interpretation of (\ref{deI}) was made in
\cite{SLP,ST} by postulating the microscopic expressions for $q$ and $f$. The
first of those is very simple
\begin{equation}
q[\rho]=\int d\mathbf{x}Y_{\mathbf{x}}\rho_{\mathbf{x}}, \label{qI}%
\end{equation}
where $\rho_{\mathbf{x}}$ is the number density, while $Y_{\mathbf{x}}$ is
some function of $\mathbf{x}$, whose integral with the spherically symmetric
distribution equals 0. The $f$ was taken in the form $f=-\partial_{q}W$, where
$W$ is a certain functional of $\rho_{\mathbf{x}}$ only \cite{ST}. Four
parameters of this functional: central density, compression modulus, central
energy density, and a quantity controlling the diffuseness of the nuclear
surface, were treated phenomenologically. The expressions for $W[\rho]$ and
$q[\rho]$ were used in the variational problem \
\begin{equation}
\frac{\delta}{\delta\rho_{\mathbf{x}}}\left(  W[\rho]+\Lambda q[\rho]-\mu\int
d\mathbf{y}\rho_{\mathbf{y}}\right)  =0, \label{ve}%
\end{equation}
whose solution provides the number density $\rho(\mathbf{x},\Lambda,\mu)$ for
the nucleus with the particle number $A=\int d\mathbf{x}\rho(\mathbf{x}%
,\Lambda,\mu)$ and deformation $q=\int d\mathbf{x}Y_{\mathbf{x}}%
\rho(\mathbf{x},\Lambda,\mu)$. Although Eq. (\ref{ve}) closely resembles the
equation for electronic density in the \emph{ground} \emph{state} of a diatom
molecule at frozen internuclear distance, with $W[\rho]$ being the analog of
the sum of the electronic kinetic energy in Thomas-Fermi approximation and the
interelectronic interaction, and $\Lambda Y_{\mathbf{x}}$ being the similitude
of the electron-nuclei potential, the authors \cite{ST} insist that $W$ and
$\rho_{\mathbf{y}}$ in (\ref{ve}) are the total energy and number density
\emph{averaged over} \emph{a group of levels} of a nucleus with fixed $A$ and
over $A$.

The $W$ in the expression $f=-\partial_{q}W$ is known as the deformation
potential. From the above we see that in Ref. \cite{ST} it is identified in
fact with the internal (or thermal) energy $U$. This contradicts to many other
works (see \cite{RKNA}, and references therein), interpreting $W$ as
Helmholtz's free energy $F$ either without proof or by assuming that the
process of deformation of a hot nucleus is similar to compression or
decompression of a gas embedded in a heat bath with a fixed temperature and
occurring so slowly that the temperature of the gas always coincides with that
of the heat bath. Evidently this analogy is misleading, because the
temperature of the nucleus in the process of deformation is increasing.

For cold nuclei the $W$ is identified with the \emph{ground} \emph{state}
energy of a nucleus with fixed $A$ and $q$. This allows one to obtain it by
minimizing the expectation $\left\langle \Psi_{g}\right\vert \mathcal{H}%
\left\vert \Psi_{g}\right\rangle $ of the total Hamiltonian $\mathcal{H}$ with
a state $\left\vert \Psi_{g}\right\rangle $ subject to the condition
$\left\langle \Psi_{g}\right\vert Q\left\vert \Psi_{g}\right\rangle =q$. Here,
$Q$ is the operator of collective coordinate%
\begin{equation}
Q=\int d\mathbf{x}Y_{\mathbf{x}}n_{\mathbf{x}}, \label{Qc}%
\end{equation}
where $n_{\mathbf{x}}=\psi_{\mathbf{x}}^{\ast}\psi_{\mathbf{x}}$ is the number
density operator and $Y_{\mathbf{x}}$ is the same as in (\ref{qI}). To
simplify this variational problem, the $\left\vert \Psi_{g}\right\rangle $ is
replaced by a \emph{single} Slater determinant $\left\vert q\right\rangle $ of
single-'particle' wave functions and, to compensate for such a severe
restriction of the Hilbert space, the bare internucleon interaction in the
expression for $\left\langle q\right\vert \mathcal{H}\left\vert q\right\rangle
$ is replaced by the effective force. This leads to the so-called
self-consistent mean-field (SCMF) models. In the early stage \cite{QF}, most
SCMF calculations have used effective forces of the Skyrme type \cite{Skyrme}
revived by Vautherin and Brink \cite{VB,V}. Nowadays \cite{BHR}, the three
most widely used variants of SCMF's are based on a Skyrme energy functional, a
Gogny force \cite{Gogny}, and a relativistic mean-field Lagrangian \cite{SW}.

To justify description of many-particle systems with strong interparticle
interactions with the aid of the SCMF model it is necessary to assume
\cite{Negele} that only the two-particle scatterings, influenced by the media,
are essential, while the three- and more particle collisions are negligible.
Moreover, the coalescence of few nucleons into a cluster is prohibited. It is
hardly possible to validate those assumptions for real nuclei. Therefore we
trust in the SCMF models simply because with several parameters in the
effective force, adjusted for a few reference nuclei, they reproduce very good
the binding energies and densities of many other nuclei \cite{BHR}.

Encouraged by good performance of the SCMF theory in studies of these
quantities, Flocard \textit{et al} \cite{Flocard} apply it with a reasonable
success to explore the two-humped fission barrier in $^{240}$Pu addressed
before within the shell-correction approach \cite{BDJPSW}. Excellent
reproduction of fission barriers for $^{240}$Pu and other acitinide nuclei,
obtained later with the Skyrme force named SkM$^{\ast}$ \cite{BGH}, seems
mysterious, because the identification of the fission path with a sequence of
\emph{ground} \emph{states} at fixed $q$ is in apparent conflict with the
\emph{compound nucleus} model of fission, used to deduce the fission barriers
from experimental excitation functions.

It is well known that for the inertial parameter $B$ in Eq. (\ref{deI}) the
hydrodynamic formula \cite{HW} is used by far more often than any other. This
poses the problem of the microscopic interpretation of this formula, to the
solution of which most close is the adiabatic time-dependent Hartree-Fock
(ATDHF) theory \cite{Villars,BV,BBIG}. A nonstationary state vector of the
nucleus $\left\vert \dot{q}q\right\rangle =e^{-i\chi}\left\vert q\right\rangle
$ is generated in this theory by appending to the Slater determinant
$\left\vert q\right\rangle $ a phase factor $e^{-i\chi}$, where $\chi$ is a
small, time-even, nonlocal one-body operator. Assuming that $\chi$ has the
form $\chi=m\dot{q}\widetilde{\phi}$, with $\widetilde{\phi}=\int
d\mathbf{x}\widetilde{\phi}_{\mathbf{x}}n_{\mathbf{x}}$, one expands the
energy $\left\langle \dot{q}q\right\vert \mathcal{H}\left\vert \dot
{q}q\right\rangle $ in powers of $\chi$, stops at second order, and dropping
the $-\dot{q}m\left\langle q\right\vert i\left[  \mathcal{H},\widetilde{\phi
}\right]  \left\vert q\right\rangle $ term from time reversal symmetry, one
gets%
\[
\left\langle \dot{q}q\right\vert \mathcal{H}\left\vert \dot{q}q\right\rangle
=\left\langle q\right\vert \mathcal{H}\left\vert q\right\rangle +\frac{1}%
{2}B\dot{q}^{2},
\]
where
\[
B=m^{2}\left\langle q\right\vert i\left[  i\left[  \mathcal{H},\widetilde
{\phi}\right]  ,\widetilde{\phi}\right]  \left\vert q\right\rangle =m\int
d\mathbf{x}\left\langle q\right\vert n_{\mathbf{x}}\left\vert q\right\rangle
\left(  \nabla\widetilde{\phi}_{\mathbf{x}}\right)  ^{2}%
\]
has the form of the inertia parameter of an abstract fluid with a mass density
$m\left\langle q\right\vert n_{\mathbf{x}}\left\vert q\right\rangle $ and a
velocity potential $\widetilde{\phi}_{\mathbf{x}}$. It should be stressed,
however, that the expression $\chi=m\dot{q}\widetilde{\phi}$ is only a rough
estimate of the solution of very involved equations for $\chi$. Moreover, the
velocity potential $\widetilde{\phi}_{\mathbf{x}}$ is not proved to satisfy
the Laplace equation and it is chosen heuristically.

Turning to microscopic expressions for friction $\gamma$, we remark that for
quite a long time \cite{H,KHR,KR,K78,YHS,Aleshin99,Aleshin03,Aleshin05} they
were extracted from the linear response function of a thermally equilibrated
Fermi gas confined by the external forces on the small variations of those
forces. And only recently \cite{Aleshin07} it has been realized that in the
bulk of such a gas, there will be no \emph{mass} \emph{flow}, $\mathbf{j}%
_{\mathbf{x}}$, defined as the expectation of the momentum density operator%
\begin{equation}
\mathbf{p}_{\mathbf{x}}=-\left(  i/2\right)  \left[  \psi_{\mathbf{x}}^{\ast
}\nabla\psi_{\mathbf{x}}-\left(  \nabla\psi_{\mathbf{x}}^{\ast}\right)
\psi_{\mathbf{x}}\right]  \label{px}%
\end{equation}
with the density matrix of the system. Therefore, the collective motion, hence
the dissipation of collective energy in such a gas will be confined to the
area of the nuclear surface. Having recognized this fact we have constructed
the Rayleigh dissipation function $\Phi$ from the entropy production rate for
the canonical ensemble $\rho_{\mathrm{l}}$ subject to the
constraint\ $\mathrm{tr}\left(  \mathbf{p}_{\mathbf{x}}\rho_{\mathrm{l}%
}\right)  =\mathbf{j}_{\mathbf{x}}$, where $\mathbf{j}_{\mathbf{x}}\neq0$ over
the entire nucleus. Although the final expressions for $\gamma$ in
\cite{Aleshin07} look reasonable, their derivation is far from convincing.
First, the $\rho_{\mathrm{l}}$ was not constrained from the beginning by the
condition $\mathrm{tr}\left(  Q\rho_{\mathrm{l}}\right)  =q$, instead of which
the deformation $q$ was introduced in $\Phi$ only in the final stage of
derivation in an intuitive way. Second, having postulated for the mass flow
$\mathbf{j}_{\mathbf{x}}$ the hydrodynamic-like formula: $\mathbf{j}%
_{\mathbf{x}}=m\rho_{\mathbf{x}}\mathbf{u}_{\mathbf{x}}$, we failed to reveal
the microscopic content of the 'fluid velocity' $\mathbf{u}_{\mathbf{x}}$,
which was treated instead according to the model of Hill and Wheeler \cite{HW}.

The main objective of this paper is the derivation of Eq. (\ref{deI}) from
first principles and removing in this way all the obscurities in microscopic
interpretations of $f$, $B$, and $\gamma$, mentioned above. Skipping for
simplicity the spin and isospin indices and letting $\hbar=1$, as done already
in (\ref{px}), the total energy and number operators are written as
\begin{equation}
\mathcal{H}=-\frac{1}{2m}\int d\mathbf{x}\psi_{\mathbf{x}}^{\ast}\nabla
^{2}\psi_{\mathbf{x}}+\frac{1}{2}\int d\mathbf{x}d\mathbf{y}v(|\mathbf{x}%
-\mathbf{y}|)\psi_{\mathbf{x}}^{\ast}\psi_{\mathbf{y}}^{\ast}\psi_{\mathbf{y}%
}\psi_{\mathbf{x}}, \label{Hcal}%
\end{equation}%
\begin{equation}
\mathcal{N}=\int d\mathbf{x}\psi_{\mathbf{x}}^{\ast}\psi_{\mathbf{x}%
},\ \label{Nc}%
\end{equation}
in terms of the field operators $\psi_{\mathbf{x}}^{\ast}$, $\psi_{\mathbf{x}%
}$, which satisfy the anticommutation relations
\begin{equation}
\psi_{\mathbf{x}}^{\ast}\psi_{\mathbf{y}}+\psi_{\mathbf{y}}\psi_{\mathbf{x}%
}^{\ast}=\delta(\mathbf{x}-\mathbf{y}),\text{ \ }\psi_{\mathbf{x}}%
\psi_{\mathbf{y}}+\psi_{\mathbf{y}}\psi_{\mathbf{x}}=0,\text{ \ }%
\psi_{\mathbf{x}}^{\ast}\psi_{\mathbf{y}}^{\ast}+\psi_{\mathbf{y}}^{\ast}%
\psi_{\mathbf{x}}^{\ast}=0, \label{ar}%
\end{equation}
and a 'bare' NN-interaction $v(|\mathbf{x}-\mathbf{y}|)$. In order to furnish
the formal description of macroscopic motion in a nucleus with total particle
number $A$ and total energy $E_{\mathrm{x}}$, Eqs. (\ref{Hcal}), (\ref{Nc})
for $\mathcal{H}$ and $\mathcal{N}$ must be supplemented by explicit
expressions in terms of $\psi_{\mathbf{x}}^{\ast}$, $\psi_{\mathbf{x}}$ for
the set $\left(  P_{1},P_{2},P_{3}\right)  =\left(  Q,P,\mathcal{M}\right)  $
of coordinate, momentum, and inertia of this motion. Denoting as $p_{n}(T)$
the expectation values of $P_{n}$ with yet unknown time-dependent density
matrix, we postulate that those $p_{n}(T)$ are very slow functions of $T$. In
formal terms this signifies that the time axis $T$ can be broken up into
succession of intervals $\left[  T_{i},T_{i}+\tau_{\mathrm{q}}\right]  $,
where $\tau_{\mathrm{q}}$ is small enough for the expressions%
\begin{equation}
p_{n}(T_{i}+t)-p_{n}(T_{i})=t\partial p_{n}(T_{i})/\partial T_{i} \label{tq}%
\end{equation}
to hold for all $t\subset$ $\left[  0,\tau_{\mathrm{q}}\right]  $, but yet
sufficiently large compared to the internal characteristic times to justify
description of the nucleus within the corresponding interval of $T$ by the
density matrix $\rho_{\mathrm{q}}$, which maximizes the entropy $S_{\mathrm{q}%
}=-\mathrm{tr}\left(  \rho_{\mathrm{q}}\ln\rho_{\mathrm{q}}\right)  $ under
the following constraints
\begin{equation}
\mathrm{tr}\left(  \mathcal{N}\rho_{\mathrm{q}}\right)  =A,\text{
\ \ \ \ \ }\mathrm{tr}\left(  \mathcal{H}\rho_{\mathrm{q}}\right)
=E_{\mathrm{x}},\text{\ }\ \text{\ \ \ \ \ }\mathrm{tr}\left(  P_{n}%
\rho_{\mathrm{q}}\right)  =p_{n}(T_{i}). \label{NHP}%
\end{equation}
According to this definition suggested by Jaynes \cite{Jaynes}, the
$\rho_{\mathrm{q}}$ does not depend on the local time $t=T-T_{i}$ of the stage
in question. The $t$ dependence is attached to the observables: $X_{t}%
=e^{it\mathcal{H}}Xe^{-it\mathcal{H}}$, where $X$ are functionals of
$\psi_{\mathbf{x}}^{\ast}$, $\psi_{\mathbf{x}}$. In particular the values of
$p_{n}\left(  T\right)  $ at the beginning of the next to $i$ stage are given
by $p_{n}(T_{i+1})=\mathrm{tr}\left(  P_{n,\tau_{\mathrm{q}}}\rho_{\mathrm{q}%
}\right)  $, which solves, in principle at least, the whole problem of
collective or macroscopic motion.

The operator $Q$ is given in (\ref{Qc}). The operators $P$ and $\mathcal{M}$
are expected to be linear functionals of the momentum density $\mathbf{p}%
_{\mathbf{x}}$ and number density $n_{\mathbf{x}}$, respectively, subject to
the conditions
\[
\mathrm{tr}\left(  P\rho_{\mathrm{q}}\right)  =\dot{q}\mathrm{tr}\left(
\mathcal{M}\rho_{\mathrm{q}}\right)  ,\ \ \ \ \ \ \ \mathrm{tr}\left(  \left[
Q,P\right]  \rho_{\mathrm{q}}\right)  =i,
\]
with $\dot{q}=\partial q(T_{i})/\partial T_{i}$ \cite{Villars}. But those
conditions alone do not determine the weighting functions in $P$ and
$\mathcal{M}$ in a unique way. Moreover, the equation $\mathrm{tr}\left(
P\rho_{\mathrm{q}}\right)  =\dot{q}\mathrm{tr}\left(  \mathcal{M}%
\rho_{\mathrm{q}}\right)  $ shows that, given $\dot{q}$, there are
correlations between $\mathrm{tr}\left(  P\rho_{\mathrm{q}}\right)  $ and
$\mathrm{tr}\left(  \mathcal{M}\rho_{\mathrm{q}}\right)  $, whereas the
constraining conditions in Jaynes's definition of $\rho_{\mathrm{q}}$ are
mutually independent. As shown in section 2, all these difficulties in
constructing $\rho_{\mathrm{q}}$ are avoided by imposing on $\rho_{\mathrm{q}%
}$ the following restriction
\[
\rho_{\mathrm{q}}=\rho_{g}\left[  \psi^{\prime\ast},\psi^{\prime}\right]
,\text{ \ \ \ \ }\psi_{\mathbf{x}}^{\prime}\equiv e^{-i\chi_{\mathbf{x}}}%
\psi_{\mathbf{x}},\text{ \ \ \ \ }\psi_{\mathbf{x}}^{\prime\ast}\equiv
e^{i\chi_{\mathbf{x}}}\psi_{\mathbf{x}}^{\ast},
\]
where the $\rho_{g}$ is the canonical density matrix, which satisfies the
conditions $\mathrm{tr}\left(  Q\rho_{g}\right)  =q$, $\mathrm{tr}\left(
\mathcal{N}\rho_{g}\right)  =A$ and depends on yet unspecified inverted
temperature $\beta$, whereas $\chi_{\mathbf{x}}$ is yet unknown real function
of the nucleon position $\mathbf{x}$. In order to work out the equations for
$\chi_{\mathbf{x}}$ and $\beta$, we use the continuity equation $\partial
_{t}\rho_{\mathbf{x},t}=-m^{-1}\nabla\mathbf{j}_{\mathbf{x},t}$ and assume
that macroscopic variables $p_{n,t}=\mathrm{tr}\left(  P_{n,t}\rho
_{\mathrm{q}}\right)  $ change with time much slower than the number and
momentum densities, $\rho_{\mathbf{x},t}$ and $\mathbf{j}_{\mathbf{x},t}$,
through which those $p_{n,t}$ are defined, the argument being that after
spatial integrations, invoked in those definitions only the slow components
$\overline{\rho}_{\mathbf{x},t}$, $\overline{\mathbf{j}}_{\mathbf{x},t}$ of
$\rho_{\mathbf{x},t}$, $\mathbf{j}_{\mathbf{x},t}$ survive. Having postulated
explicit expressions for $\overline{\rho}_{\mathbf{x},t}$, $\overline
{\mathbf{j}}_{\mathbf{x},t}$, we complete the definition of $\chi_{\mathbf{x}%
}$ and $\beta$, the net result being that the $\rho_{\mathrm{q}}$ is now
completely defined, which allows one to use Eq. (\ref{NHP}) for the definition
of the weighting functions in $P$ and $\mathcal{M}$ in a unique way.

In section 3 we utilize another, besides (\ref{tq}), an implication of the
adiabaticity hypothesis to get the closed-form solution of the dynamic
equation for the expectation $\mathrm{tr}\left(  K_{t}\rho_{\mathrm{q}%
}\right)  $ of $K_{t}=e^{it\mathcal{H}}Ke^{-it\mathcal{H}}$, $K=P-$ $\dot
{q}\mathcal{M}$, and use this solution to express the quotient $q^{\prime
\prime}=\left[  \dot{q}\left(  T_{i}+\tau_{\mathrm{q}}\right)  -\dot{q}\left(
T_{i}\right)  \right]  /\tau_{\mathrm{q}}$ as a function of the quotient
$q^{\prime}=\left[  q\left(  T_{i}+\tau_{\mathrm{q}}\right)  -q\left(
T_{i}\right)  \right]  /\tau_{\mathrm{q}}$ and $q=q\left(  T_{i}\right)  $.
After replacement of those quotients with the derivatives, this expression
takes the form of equation (\ref{deI}) for $q(T)$, which permits us to get the
many-body expressions for $B$, $f$, and $\gamma$.

In section 4, we use the results of Refs.
\cite{BalianDominicis,UMT,Neumann,Mermin} to construct the self-confined Fermi
gas (SCFG) model, which is a natural extension of the self-consistent mean
field model of ground state nuclei with fixed shape, to hot nuclei. The
expressions for $B$, $f$, and $\gamma$ obtained for the self-confined Fermi
gas from our general theory of nuclear collective motion are compared with the
previous models of those quantities. In section 5 we give a short summary of
our main conclusions and elucidate the physical nature of collective motion in
hot nuclei.

\section{The quasiequilibrium density matrix}

\subsection{General expressions for $\rho_{\mathrm{q}}$}

In this section we define the quasiequilibrium density matrix $\rho
_{\mathrm{q}}$ of the system on a \emph{separate stage} of collective motion,
which lasts between $T=T_{i}$ and $T=T_{i}+\tau_{\mathrm{q}}$. This
$\rho_{\mathrm{q}}$ does not depend on the local time $t=T-T_{i}$, while all
observables depend on $t$ according to the low
\begin{equation}
X_{t}=e^{i\mathcal{H}t}Xe^{-i\mathcal{H}t}, \label{Hr}%
\end{equation}
where $X\equiv$ $\left.  X_{t}\right\vert _{t=0}$. The values of $q(T_{i})$
and $\partial q(T_{i})/\partial T_{i}$ are denoted as $q$ and $\dot{q}$,
respectively. The total number and the total energy of the system are $A$ and
$E_{\mathrm{x}}$.

Given $q$, the canonical density matrix of the nucleus reads
\begin{equation}
\rho_{g}=Z_{g}^{-1}e^{-\beta(\mathcal{H}-\mu_{g}\mathcal{N}+\Lambda_{g}Q)},
\label{rhoeq}%
\end{equation}%
\begin{equation}
Z_{g}=\mathrm{tr}e^{-\beta(\mathcal{H}-\mu_{g}\mathcal{N}+\Lambda_{g}Q)},
\label{Zg}%
\end{equation}
where the $\mu_{g}$ and $\Lambda_{g}$ are functions of $\beta$, $A$, $q$,
determined from the equations
\begin{equation}
\mathrm{tr}\left(  \mathcal{N}e^{-\beta(\mathcal{H}-\mu\mathcal{N}+\Lambda
Q)}\right)  =A\mathrm{tr}e^{-\beta(\mathcal{H}-\mu\mathcal{N}+\Lambda
Q)},\text{ \ \ \ }\mathrm{tr}\left(  Qe^{-\beta(\mathcal{H}-\mu\mathcal{N}%
+\Lambda Q)}\right)  =q\mathrm{tr}e^{-\beta(\mathcal{H}-\mu\mathcal{N}+\Lambda
Q)} \label{NQ}%
\end{equation}
with respect to $\mu$, $\Lambda$, while the inverted temperature $\beta$ will
be specified later. Putting into (\ref{NQ}) the definitions (\ref{Nc}),
(\ref{Qc}) of $\mathcal{N}$ and $Q$, respectively, and introducing the number
density
\begin{equation}
\rho_{\mathbf{x}}\equiv\rho_{A,\beta}(\mathbf{x},q)\equiv\rho(\mathbf{x}%
;\beta,\mu_{g},\Lambda_{g})=\mathrm{tr}\left(  n_{\mathbf{x}}\rho_{g}\right)
, \label{rng}%
\end{equation}
we rewrite the equations for $\mu_{g}$ and $\Lambda_{g}$ in the form
\begin{equation}
\int d\mathbf{x}\rho(\mathbf{x};\beta,\mu,\Lambda)=A,\text{ \ \ \ \ \ \ }\int
d\mathbf{x}Y_{\mathbf{x}}\rho(\mathbf{x};\beta,\mu,\Lambda)=q, \label{mL}%
\end{equation}
from which it follows that
\begin{equation}
\int d\mathbf{x}\rho_{\mathbf{x}}=A, \label{Ir=A}%
\end{equation}%
\begin{equation}
\int d\mathbf{x}Y_{\mathbf{x}}\rho_{\mathbf{x}}=q. \label{IYr=q}%
\end{equation}

The mass flow in $\rho_{g}$ equals 0: \
\begin{equation}
\mathrm{tr}\left(  \mathbf{p}_{\mathbf{x}}\rho_{g}\right)  =0, \label{pxg}%
\end{equation}
because $\rho_{g}$, being a function of even operator $\mathcal{H}-\mu
_{g}\mathcal{N}+\Lambda_{g}Q$, is itself even under a time-reversal
transformation, while $\mathbf{p}_{\mathbf{x}}$ is odd. In order to find the
general form of $\rho_{\mathrm{q}}$ with $\mathrm{tr}\left(  \mathbf{p}%
_{\mathbf{x}}\rho_{\mathrm{q}}\right)  \neq0$, we use the analogy with the
classical fluid with the velocity $\overline
{\mathbf{\mathbf{\mathbf{\mathbf{\mathbf{u}}}}}}_{\mathbf{x}}\neq0$. According
to Mori \cite{Mori} the distribution function of such a fluid, consisting of
particles of mass $m$ with coordinates $\mathbf{x}_{i}$ and
momenta$\ \mathbf{p}_{i}$, is obtained from that of the fluid with
$\overline{\mathbf{\mathbf{\mathbf{\mathbf{\mathbf{u}}}}}}_{\mathbf{x}}=0$ by
the substitutions%
\begin{equation}
\mathbf{x}_{i}\rightarrow\mathbf{x}_{i}^{\prime}=\mathbf{x}_{i},\text{
\ \ \ \ \ }\mathbf{p}_{i}\rightarrow\mathbf{p}_{i}^{\prime}=\mathbf{p}%
_{i}-m\overline{\mathbf{\mathbf{\mathbf{\mathbf{\mathbf{u}}}}}}_{\mathbf{x}}.
\label{x'p'}%
\end{equation}
Keeping in mind that (\ref{x'p'}) is the canonical transformation, because
$\mathbf{x}_{i}^{\prime}$, $\mathbf{p}_{i}^{\prime}$ obey the same Poisson
bracket as $\mathbf{x}_{i}$, $\mathbf{p}_{i}$, and that in quantum systems of
identical particles the role of $\mathbf{x}_{i}$ and $\mathbf{p}_{i}$ is taken
over by $\psi_{\mathbf{x}}$, $\psi_{\mathbf{x}}^{\ast}$, we therefore
\emph{define} the density matrix $\rho_{\mathrm{q}}$ with $\mathrm{tr}\left(
\mathbf{p}_{\mathbf{x}}\rho_{\mathrm{q}}\right)  \neq0$ as follows
\[
\rho_{\mathrm{q}}=\rho_{g}^{\prime},\text{ \ \ \ \ \ \ \ \ \ }\rho_{g}%
^{\prime}\equiv\rho_{g}\left[  \psi^{\prime\ast},\psi^{\prime}\right]  ,
\]
where $\psi_{\mathbf{x}}^{\prime}$, $\psi_{\mathbf{x}}^{\prime\ast}$ must
satisfy the canonical anticommutation relations (\ref{ar}) and give rise to
the primed densities
\begin{equation}
n_{\mathbf{x}}^{\prime}\equiv\psi_{\mathbf{x}}^{\prime\ast}\psi_{\mathbf{x}%
}^{\prime},\text{ \ \ \ }\mathbf{p}_{\mathbf{x}}^{\prime}\equiv-\frac{i}%
{2}\left[  \psi_{\mathbf{x}}^{\prime\ast}\nabla\psi_{\mathbf{x}}^{\prime
}-(\nabla\psi_{\mathbf{x}}^{\prime\ast})\psi_{\mathbf{x}}^{\prime}\right]  ,
\label{n'xp'x}%
\end{equation}
obeying the conditions
\begin{equation}
n_{\mathbf{x}}^{\prime}=n_{\mathbf{x}},\text{ \ \ \ \ \ \ }\mathbf{p}%
_{\mathbf{x}}^{\prime}=\mathbf{p}_{\mathbf{x}}-mn_{\mathbf{x}}%
\mathbf{\mathbf{\mathbf{\mathbf{\mathbf{u}}}}}_{\mathbf{x}}, \label{n'n}%
\end{equation}
with $m$ being the nucleon mass and
$\mathbf{\mathbf{\mathbf{\mathbf{\mathbf{u}}}}}_{\mathbf{x}}$ an yet unknown
c-number vector function. Eqs. (\ref{n'n}) have exactly the same form as the
relationships, following from Eqs. (\ref{x'p'}), between the corresponding
densities in the classical fluid.

It is readily seen that the required $\psi_{\mathbf{x}}^{\prime}$,
$\psi_{\mathbf{x}}^{\prime\ast}$ have the form
\begin{equation}
\psi_{\mathbf{x}}^{\prime}=e^{-i\chi_{\mathbf{x}}}\psi_{\mathbf{x}},\text{
\ \ \ }\psi_{\mathbf{x}}^{\prime\ast}=e^{i\chi_{\mathbf{x}}}\psi_{\mathbf{x}%
}^{\ast}, \label{t1}%
\end{equation}
where $\chi_{\mathbf{x}}$ is a real function remaining to be determined.
Really, the fulfillment of anticommutation relations and of the first
condition in (\ref{n'n}) is evident, while substitution of (\ref{t1}) into
$\mathbf{p}_{\mathbf{x}}^{\prime}$ defined in (\ref{n'xp'x}) leads to the
second condition in (\ref{n'n}) with
\begin{equation}
\mathbf{\mathbf{\mathbf{\mathbf{\mathbf{u}}}}}_{\mathbf{x}}=m^{-1}\nabla
\chi_{\mathbf{x}}. \label{pot0}%
\end{equation}

The trace of any functional $\mathcal{A}$ of the nucleon field operators is
\begin{equation}
\mathrm{tr}\mathcal{A}=\sum_{N=0}^{\infty}\frac{1}{N!}\int d\mathbf{x}%
_{1},d\mathbf{x}_{2...}d\mathbf{x}_{N}\left\langle 0\right\vert \psi
_{\mathbf{x}_{1}}\psi_{\mathbf{x}_{2}}...\psi_{\mathbf{x}_{N}}\mathcal{A}%
\psi_{\mathbf{x}_{N}}^{\ast}...\psi_{\mathbf{x}_{2}}^{\ast}\psi_{\mathbf{x}%
_{1}}^{\ast}\left\vert 0\right\rangle , \label{tr1}%
\end{equation}
$\left\vert 0\right\rangle $ being the state with no particles. By definition,
the primed trace $\mathrm{tr}^{\prime}\mathcal{A}$ is obtained from
$\mathrm{tr}\mathcal{A}$ by putting primes on those $\psi$ and $\psi^{\ast}$,
which explicitly appear in (\ref{tr1}), but not on those hidden in
$\mathcal{A}$. From (\ref{t1}) then follows the formula $\mathrm{tr}^{\prime
}\mathcal{A}=\mathrm{tr}\mathcal{A}$ , which leads at $\mathcal{A}%
=X\rho_{\mathrm{q}}$ with accounting for $\rho_{\mathrm{q}}=\rho_{g}^{\prime}$
to the useful relationship
\begin{equation}
\mathrm{tr}\left(  X\rho_{\mathrm{q}}\right)  =\mathrm{tr}^{\prime}\left(
X\rho_{g}^{\prime}\right)  . \label{trtr'}%
\end{equation}
As a first application of (\ref{trtr'}) consider the mass flow in
$\rho_{\mathrm{q}}$:
\begin{equation}
\mathbf{j}_{\mathbf{x}}=\mathrm{tr}\left(  \mathbf{p}_{\mathbf{x}}%
\rho_{\mathrm{q}}\right)  . \label{jx}%
\end{equation}
From (\ref{trtr'}) and the expression
\begin{equation}
\mathbf{p}_{\mathbf{x}}=\mathbf{p}_{\mathbf{x}}^{\prime}+mn_{\mathbf{x}%
}^{\prime}\mathbf{\mathbf{\mathbf{\mathbf{\mathbf{\mathbf{u}}}}}}_{\mathbf{x}%
}, \label{pp'}%
\end{equation}
which follows from (\ref{n'n}), one obtains
\begin{equation}
\mathbf{j}_{\mathbf{x}}=\mathrm{tr}^{\prime}\left[  (\mathbf{p}_{\mathbf{x}%
}^{\prime}+mn_{\mathbf{x}}^{\prime}%
\mathbf{\mathbf{\mathbf{\mathbf{\mathbf{\mathbf{u}}}}}}_{\mathbf{x}})\rho
_{g}^{\prime}\right]  . \label{j0}%
\end{equation}
By definition, the primed counterpart $X^{\prime}$ of any operator $X$,
including the trace, is expressed in terms of $\psi_{\mathbf{x}}^{\prime}$,
$\psi_{\mathbf{x}}^{\prime\ast}$ in the same way that the corresponding
unprimed operator $X$ is expressed in terms of $\psi_{\mathbf{x}}$,
$\psi_{\mathbf{x}}^{\ast}$. Since, in addition, $\psi_{\mathbf{x}}^{\prime}$,
$\psi_{\mathbf{x}}^{\prime\ast}$ obey the same anticommutator relations as
$\psi_{\mathbf{x}}$, $\psi_{\mathbf{x}}^{\ast}$, one can drop all primes in
(\ref{j0}). Then, using (\ref{pxg}), (\ref{rng}) yields
\begin{equation}
\mathbf{j}_{\mathbf{x}}=m\rho_{A,\beta}(\mathbf{x}%
,q)\mathbf{\mathbf{\mathbf{\mathbf{\mathbf{\mathbf{u}}}}}}_{\mathbf{x}}.
\label{jxq}%
\end{equation}

Let us now employ Eq. (\ref{trtr'}) at $X=-\ln\rho_{\mathrm{q}}$ to calculate
the entropy $S_{\mathrm{q}}=-\mathrm{tr}\left(  \rho_{\mathrm{q}}\ln
\rho_{\mathrm{q}}\right)  $ in the $\rho_{\mathrm{q}}$ ensemble. Use of
$\rho_{\mathrm{q}}=\rho_{g}^{\prime}$ leads to $S_{\mathrm{q}}=-\mathrm{tr}%
^{\prime}\left(  \rho_{g}^{\prime}\ln\rho_{g}^{\prime}\right)  $, or on
removal of all primes,
\[
S_{\mathrm{q}}=S,
\]
where $S=-\mathrm{tr}\left(  \rho_{g}\ln\rho_{g}\right)  $ is the entropy of
the $\rho_{g}$ ensemble.

In passing we note that $\psi_{\mathbf{x}}^{\prime}$ and $\psi_{\mathbf{x}%
}^{\prime\ast}$ given in (\ref{t1}) can be put into the form
\begin{equation}
\psi_{\mathbf{x}}^{\prime}=e^{-i\chi}\psi_{\mathbf{x}}e^{i\chi},\text{
\ \ \ \ \ }\psi_{\mathbf{x}}^{\prime\ast}=e^{-i\chi}\psi_{\mathbf{x}}^{\ast
}e^{i\chi}, \label{Br}%
\end{equation}
where
\begin{equation}
\chi=\int d\mathbf{x}\chi_{\mathbf{x}}\psi_{\mathbf{x}}^{\ast}\psi
_{\mathbf{x}}=\int d\mathbf{x}\chi_{\mathbf{x}}n_{\mathbf{x}} \label{po}%
\end{equation}
will be called the phase operator. This is proven by application of the
formula
\begin{equation}
e^{i\mathcal{A}}Xe^{-i\mathcal{A}}=X+i\left[  \mathcal{A},X\right]
+\frac{i^{2}}{2!}\left[  \mathcal{A},\left[  \mathcal{A},X\right]  \right]
+... \label{eXe}%
\end{equation}
at $\mathcal{A}=-\chi$ and $X=\psi_{\mathbf{x}}$, $\psi_{\mathbf{x}}^{\ast}$
and the expressions
\[
\left[  \chi,\psi_{\mathbf{x}}\right]  =-\chi_{\mathbf{x}}\psi_{\mathbf{x}%
},\text{ \ \ \ \ }\left[  \chi,\psi_{\mathbf{x}}^{\ast}\right]  =\chi
_{\mathbf{x}}\psi_{\mathbf{x}}^{\ast}.
\]
Eq. (\ref{Br}) allows the definition $X^{\prime}\equiv X\left[  \psi
^{\prime\ast},\psi^{\prime}\right]  $ of any primed operator to be written as
\begin{equation}
X^{\prime}=e^{-i\chi}Xe^{i\chi}. \label{Xp}%
\end{equation}
For $X=\rho_{g}$ on remarking that $X^{\prime}=\rho_{g}^{\prime}%
=\rho_{\mathrm{q}}$, Eq. (\ref{Xp}) becomes%
\begin{equation}
\rho_{\mathrm{q}}=e^{-i\chi}\rho_{g}e^{i\chi}. \label{rqj}%
\end{equation}
This formula for $\rho_{\mathrm{q}}$ is analogous to the definition of the
density matrix of the time dependent Hartree-Fock formalism \cite{BV,BBIG}, in
which, however, the operator analogous to our $\chi$, Eq. (\ref{po}), is
nonlocal, while that analogous to $\rho_{g}$ represents a time-even Slater determinant.

\subsection{The definitions of $\chi_{\mathbf{x}}$ and $\beta$}

With Eq. (\ref{rhoeq}) for $\rho_{g}$, the definition $\rho_{\mathrm{q}}%
=\rho_{g}^{\prime}$ implies that
\begin{equation}
\rho_{\mathrm{q}}=Z_{g}^{-1}e^{-\beta\left(  \mathcal{H}^{\prime}-\mu
_{g}\mathcal{N}^{\prime}+\Lambda_{g}Q^{\prime}\right)  }, \label{rg'}%
\end{equation}
where $\mathcal{H}^{\prime}$, $\mathcal{N}^{\prime}$, and $Q^{\prime}$ are
constructed from $\psi_{\mathbf{x}}^{\prime}$ and $\psi_{\mathbf{x}}%
^{\prime\ast}$ in the same way that $\mathcal{H}$, $\mathcal{N}$, and $Q$ are
constructed from $\psi_{\mathbf{x}}$ and $\psi_{\mathbf{x}}^{\ast}$. Use of
Eqs. (\ref{Hcal}), (\ref{Nc}), (\ref{Qc}) for $\mathcal{H}$, $\mathcal{N}$,
$Q$, respectively, and Eq. (\ref{pot0}) for
$\mathbf{\mathbf{\mathbf{\mathbf{\mathbf{u}}}}}_{\mathbf{x}}$ leads to the
expressions%
\begin{equation}
\mathcal{H}^{\prime}=\mathcal{H}-\int
d\mathbf{x\mathbf{\mathbf{\mathbf{\mathbf{\mathbf{\mathbf{u}}}}}}%
}_{\mathbf{\mathbf{\mathbf{\mathbf{\mathbf{\mathbf{x}}}}}}}%
\mathbf{\mathbf{\mathbf{\mathbf{\mathbf{p}}}}}_{\mathbf{x}}+\frac{m}{2}\int
d\mathbf{xu}_{\mathbf{x}}^{2}n_{\mathbf{x}}, \label{H'}%
\end{equation}
$\mathcal{N}^{\prime}=\mathcal{N}$, $Q^{\prime}=Q$, so that (\ref{rg'})
becomes
\begin{equation}
\rho_{\mathrm{q}}=Z_{g}^{-1}\exp\left\{  -\beta\left[  \mathcal{H}+\Lambda
_{g}Q-\mu_{g}\mathcal{N}-\int
d\mathbf{x\mathbf{\mathbf{\mathbf{\mathbf{\mathbf{\mathbf{\mathbf{u}}}}}}%
}_{\mathbf{\mathbf{\mathbf{\mathbf{\mathbf{\mathbf{x}}}}}}}%
\mathbf{\mathbf{\mathbf{\mathbf{p}}}}}_{\mathbf{x}}+\frac{m}{2}\int
d\mathbf{xu}_{\mathbf{x}}^{2}n_{\mathbf{x}}\right]  \right\}  , \label{(6)}%
\end{equation}
where $\mathbf{\mathbf{\mathbf{\mathbf{\mathbf{u}}}}}_{\mathbf{x}}%
=m^{-1}\nabla\chi_{\mathbf{x}}$. In this subsection we formulate the equations
for still unknown parameters $\chi_{\mathbf{x}}$ and $\beta$ of $\rho
_{\mathrm{q}}$ by exploiting the adiabaticity of collective motion and the
continuity equation.

To embody the slowness of collective variables in their formal definition
introduce the number density $\rho_{\mathbf{x},t}$ at time $t$
\begin{equation}
\rho_{\mathbf{x},t}=\mathrm{tr}\left(  n_{\mathbf{x},t}\rho_{\mathrm{q}%
}\right)  . \label{rt}%
\end{equation}
For $t=0$, $\rho_{\mathbf{x},t}$ reduces to
\begin{equation}
\rho_{\mathbf{x},0}=\mathrm{tr}\left(  n_{\mathbf{x}}\rho_{\mathrm{q}}\right)
, \label{rx}%
\end{equation}
since $n_{\mathbf{x},0}=n_{\mathbf{x}}$. Using $n_{\mathbf{x}}=n_{\mathbf{x}%
}^{\prime}$ and the same line of reasoning as in deriving $S_{\mathrm{q}}$, we
find that $\rho_{\mathbf{x},0}=\mathrm{tr}\left(  n_{\mathbf{x}}\rho
_{g}\right)  $, or on recalling (\ref{rng})
\begin{equation}
\rho_{\mathbf{x},0}=\rho_{A,\beta}(\mathbf{x},q). \label{rx1}%
\end{equation}

Having assumed that $q_{t}\equiv q(T_{i}+t)$ varies with time much slower than
$\rho_{\mathbf{x},t}$, introduce the slow component of $\rho_{\mathbf{x},t}$
by the relation $\overline{\rho}_{\mathbf{x},t}\equiv\rho_{A,\beta}%
(\mathbf{x},q_{t})$. Some collective variables, such as $q_{t}\equiv
\mathrm{tr}\left(  Q_{t}\rho_{\mathrm{q}}\right)  $, $M_{t}\equiv
\mathrm{tr}\left(  \mathcal{M}_{t}\rho_{\mathrm{q}}\right)  $, which have the
following general form $\int d\mathbf{x}w_{\mathbf{x}}\rho_{\mathbf{x},t}$,
where $w_{\mathbf{x}}$ is a c-number function, can be put into the form
\[
\int d\mathbf{x}\rho_{\mathbf{x},t}w_{\mathbf{x}}=\int d\mathbf{x}%
\overline{\rho}_{\mathbf{x},t}w_{\mathbf{x}}+\int d\mathbf{x}\delta
\rho_{\mathbf{x},t}w_{\mathbf{x}},
\]
where $\delta\rho_{\mathbf{x},t}=\rho_{\mathbf{x},t}-\overline{\rho
}_{\mathbf{x},t}$ is a rapidly changing component of $\rho_{\mathbf{x},t}$.
Now we postulate that spatial integration in the second term completely
eliminates it, enabling one for estimating those variables with the aid of the
substitution
\begin{equation}
\rho_{\mathbf{x},t}\rightarrow\overline{\rho}_{\mathbf{x},t}\equiv
\rho_{A,\beta}(\mathbf{x},q_{t}). \label{rxt}%
\end{equation}
Likewise, the time derivatives of the collective variables of the above type
will be evaluated by replacing $\partial_{t}\rho_{\mathbf{x},t}$ with $\dot
{q}_{t}\partial_{q}\rho_{A,\beta}(\mathbf{x},q_{t})$, where $\dot{q}%
_{t}=\partial_{t}q_{t}$.

As seen from the relations
\begin{equation}
\partial_{t}\rho_{\mathbf{x},t}=\partial_{t}\mathrm{tr}\left(
e^{it\mathcal{H}}n_{\mathbf{x}}e^{-it\mathcal{H}}\rho_{\mathrm{q}}\right)
=\mathrm{tr}\left(  e^{it\mathcal{H}}i[\mathcal{H},n_{\mathbf{x}%
}]e^{-it\mathcal{H}}\rho_{\mathrm{q}}\right)  =-m^{-1}\nabla\mathrm{tr}\left(
e^{it\mathcal{H}}\mathbf{p}_{\mathbf{x}}e^{-it\mathcal{H}}\rho_{\mathrm{q}%
}\right)  , \label{dra}%
\end{equation}
relaying on (\ref{rt}), (\ref{Hr}) and the readily proven expression
\begin{equation}
i[\mathcal{H},n_{\mathbf{x}}]=-m^{-1}\nabla\mathbf{p}_{\mathbf{x}}, \label{nd}%
\end{equation}
the $\partial_{t}\rho_{\mathbf{x},t}$ can be presented as%
\begin{equation}
\partial_{t}\rho_{\mathbf{x},t}=-m^{-1}\nabla\mathbf{j}_{\mathbf{x},t},
\label{ce}%
\end{equation}
where
\begin{equation}
\mathbf{j}_{\mathbf{x},t}\equiv\mathrm{tr}\left(  \mathbf{p}_{\mathbf{x}%
,t}\rho_{\mathrm{q}}\right)  \label{jt}%
\end{equation}
is the mass flow at time $t$. For $t=0$, $\mathbf{j}_{\mathbf{x},t}$ reduces
to $\mathbf{j}_{\mathbf{x}}$ given in (\ref{jxq}), hence Eq. (\ref{ce})
becomes
\[
\left.  \partial_{t}\rho_{\mathbf{x},t}\right\vert _{t=0}=-\nabla\left[
\rho_{A,\beta}(\mathbf{x},q)\mathbf{\mathbf{\mathbf{\mathbf{\mathbf{\mathbf{u}%
}}}}}_{\mathbf{x}}\right]  .
\]
If $\left.  \partial_{t}\rho_{\mathbf{x},t}\right\vert _{t=0}$ stands under
the integral $\int d\mathbf{x...}$ over entire space with an appropriate
weighting function, then we can replace it with $\dot{q}\partial_{q}%
\rho_{A,\beta}(\mathbf{x},q)$. This yields the equality
\begin{equation}
\dot{q}\partial_{q}\rho_{\mathbf{x}}=-\nabla\left(  \rho_{\mathbf{x}%
}\mathbf{u}_{\mathbf{x}}\right)  , \label{qqe}%
\end{equation}
which we regard as an equation for $\mathbf{u}_{\mathbf{x}}$, because all
other ingredients of (\ref{qqe}) are known, except for $\beta$, which will be
determined later. Since the left side of (\ref{qqe}) is linear in $\dot{q}$
and since $\mathbf{\mathbf{\mathbf{\mathbf{\mathbf{u}}}}}_{\mathbf{x}}%
=m^{-1}\nabla\chi_{\mathbf{x}}$, (\ref{pot0}), the solution of (\ref{qqe}) has
the form
\begin{equation}
\mathbf{u}_{\mathbf{x}}=\dot{q}\mathbf{\widetilde{\mathbf{u}}}%
_{\mathbf{\mathbf{x}}}, \label{uw}%
\end{equation}%
\begin{equation}
\mathbf{\widetilde{\mathbf{u}}}_{\mathbf{\mathbf{x}}}=\nabla\widetilde{\phi
}_{\mathbf{x}}, \label{pot}%
\end{equation}
where $\widetilde{\phi}_{\mathbf{x}}=$ $\left(  \dot{q}m\right)  ^{-1}%
\chi_{\mathbf{x}}$ is the solution of the equation
\begin{equation}
\rho_{\mathbf{x}}\Delta\widetilde{\phi}_{\mathbf{x}}+\left(  \nabla
\rho_{\mathbf{x}}\right)  \nabla\widetilde{\phi}_{\mathbf{x}}=-\partial
_{q}\rho_{\mathbf{x}}, \label{icq}%
\end{equation}
which determines it up to an inessential additive constant. As $\rho
_{\mathbf{x}}$ is a function of $A$, $\beta$, $q$, the same is true concerning
$\widetilde{\phi}_{\mathbf{x}}$ and $\mathbf{\widetilde{\mathbf{u}}%
}_{\mathbf{\mathbf{x}}}$: $\widetilde{\phi}_{\mathbf{x}}\equiv\widetilde{\phi
}_{A,\beta}(\mathbf{x},q)$, $\mathbf{\widetilde{\mathbf{u}}}%
_{\mathbf{\mathbf{x}}}\equiv\mathbf{\widetilde{\mathbf{u}}}_{A,\beta
}(\mathbf{x},q)$.

In addition to using $\overline{\rho}_{\mathbf{x},t}=\rho_{A,\beta}%
(\mathbf{x},q_{t})$ instead of $\rho_{\mathbf{x},t}$ in the spatial integrals
involving $\rho_{\mathbf{x},t}$, the spatial integrals involving
$\mathbf{j}_{\mathbf{x},t}$ will be calculated with the aid of the
replacement
\begin{equation}
\text{\ }\mathbf{j}_{\mathbf{x},t}\rightarrow\overline{\mathbf{j}}%
_{\mathbf{x},t}\equiv m\rho_{A,\beta}(\mathbf{x},q_{t})\dot{q}_{t}%
\mathbf{\widetilde{\mathbf{u}}}_{A,\beta}(\mathbf{x},q_{t}). \label{uxt}%
\end{equation}
The special form of the slow component $\overline{\mathbf{j}}_{\mathbf{x},t}$
of \ $\mathbf{j}_{\mathbf{x},t}$ in (\ref{uxt}) is motivated by the expression
$\mathbf{j}_{\mathbf{x}}=m\rho_{A,\beta}(\mathbf{x},q)\dot{q}%
\mathbf{\widetilde{\mathbf{u}}}_{A,\beta}(\mathbf{x},q)$, which follows
immediately from (\ref{jxq}) and (\ref{uw}).

The knowledge of the total energy of the system imposes the following
restriction on $\rho_{\mathrm{q}}$: $\mathrm{tr}\left(  \mathcal{H}%
\rho_{\mathrm{q}}\right)  =$ $E_{\mathrm{x}}$. Accounting for (\ref{trtr'})
results in the expression $E_{\mathrm{x}}=\mathrm{tr}^{\prime}\left(
\mathcal{H}\rho_{g}^{\prime}\right)  $. Use of (\ref{n'n}) in (\ref{H'}) leads
to the formula
\begin{equation}
\mathcal{H}=\mathcal{H}^{\prime}+\int
d\mathbf{x\mathbf{\mathbf{\mathbf{\mathbf{\mathbf{\mathbf{\mathbf{u}}}}}}}%
}_{\mathbf{\mathbf{x}}}\mathbf{p_{\mathbf{\mathbf{x}}}^{\prime}%
\mathbf{\mathbf{\mathbf{+}}}}\frac{m}{2}\int d\mathbf{x}u_{\mathbf{x}}%
^{2}n_{\mathbf{x}}^{\prime}. \label{HH'}%
\end{equation}
Having inserted this expression for $\mathcal{H}$ into $E_{\mathrm{x}%
}=\mathrm{tr}^{\prime}\left(  \mathcal{H}\rho_{g}^{\prime}\right)  $, we
remove all primes and invoke (\ref{pxg}), (\ref{rng}), to find
\begin{equation}
E_{\mathrm{x}}=\frac{m}{2}\int d\mathbf{x}\rho_{\mathbf{x}}u_{\mathbf{x}}%
^{2}+U, \label{<H>q}%
\end{equation}
where
\begin{equation}
U=\mathrm{tr}\left(  \mathcal{H}\rho_{g}\right)  \label{U1}%
\end{equation}
is the internal energy of the nucleus.

Substituting (\ref{uw}) into (\ref{<H>q}), we obtain
\begin{equation}
E_{\mathrm{x}}=M\dot{q}^{2}/2+U, \label{Hq1}%
\end{equation}
where \
\begin{equation}
M=m{\int}d\mathbf{x}\rho_{\mathbf{x}}\widetilde{u}_{\mathbf{\mathbf{x}}}^{2}.
\label{Mq}%
\end{equation}
Since $U$ and $M$ are functions of $A$, $\beta$, $q$, Eq. (\ref{Hq1}) is
nothing else but an implicit equation for $\beta$, whose solution uniquely
determines $\beta$ as a function of $A$, $E_{\mathrm{x}}$, $q$, $\dot{q}$.

\subsection{The definitions of $P$ and $\mathcal{M}$}

Substituting (\ref{uw}) into (\ref{(6)}), we find for $\rho_{\mathrm{q}}$ the
final expression
\begin{equation}
\rho_{\mathrm{q}}=Z_{g}^{-1}\exp\left[  -\beta\left(  \mathcal{H}-\mu
_{g}\mathcal{N}+\Lambda_{g}Q-\dot{q}P+\frac{1}{2}\dot{q}^{2}\mathcal{M}%
\right)  \right]  , \label{frq}%
\end{equation}
where
\begin{equation}
Q\equiv\int d\mathbf{x}Y_{\mathbf{x}}n_{\mathbf{x}}, \label{Qcoll}%
\end{equation}%
\begin{equation}
P\equiv\int d\mathbf{x\mathbf{\widetilde{\mathbf{u}}}_{\mathbf{\mathbf{x}}%
}\mathbf{\mathbf{\mathbf{\mathbf{p}}}}}_{\mathbf{x}}, \label{Pcoll}%
\end{equation}%
\begin{equation}
\mathcal{M}\equiv m\int d\mathbf{x}\widetilde{u}_{\mathbf{\mathbf{x}}}%
^{2}n_{\mathbf{x}}. \label{Mcoll}%
\end{equation}
The expectation values of $Q$; $P$; $\mathcal{M}$ with $\rho_{\mathrm{q}}$ are
as follows
\begin{equation}
\mathrm{tr}\left(  Q\rho_{\mathrm{q}}\right)  =q;\text{ \ \ \ \ \ \ }%
\mathrm{tr}\left(  P\rho_{\mathrm{q}}\right)  =M\dot{q};\text{ \ \ \ \ \ }%
\mathrm{tr}\left(  \mathcal{M}\rho_{\mathrm{q}}\right)  =M. \label{trQPM}%
\end{equation}
They are obtained from (\ref{Qcoll}), (\ref{rx}), (\ref{rx1}), (\ref{IYr=q});
(\ref{Pcoll}), (\ref{jx}), (\ref{jxq}), (\ref{uw}), (\ref{Mq}); (\ref{Mcoll}),
(\ref{rx}), (\ref{rx1}), (\ref{Mq}), respectively.

The operator $Q$ defined in (\ref{Qcoll}) coincides with the collective
coordinate $Q$, introduced in (\ref{Qc}). Now we show that operators $P$ and
$\mathcal{M}$ given in (\ref{Pcoll}) and (\ref{Mcoll}) can be interpreted as
the momentum, canonically conjugated to $Q$, and the inertia of collective
motion. For this purpose, we examine the expectation values $\left\langle
K_{1}\right\rangle _{g}$, $\left\langle K_{1}\right\rangle _{\mathrm{q}}$ of
the operator
\begin{equation}
K_{1}\equiv i\left[  Q,P\right]  \label{iQP}%
\end{equation}
and relate $\left\langle P\right\rangle _{\mathrm{q}}$ with $\left\langle
\mathcal{M}\right\rangle _{\mathrm{q}}$. Here and in the following the
expectation values of an arbitrary observable $X$ in the ensembles $\rho_{g}$
and $\rho_{\mathrm{q}}$ are denoted as
\[
\left\langle X\right\rangle _{g}\equiv\mathrm{tr}(X\rho_{g}%
),\text{\ \ \ \ \ \ }\left\langle X\right\rangle _{\mathrm{q}}\equiv
\mathrm{tr}(X\rho_{\mathrm{q}}).
\]

From (\ref{iQP}), (\ref{Qcoll}), (\ref{Pcoll}) and the commutation relation
\begin{equation}
\int d\mathbf{y}f_{\mathbf{y}}i\left[  n_{\mathbf{y}},\mathbf{p}_{\mathbf{x}%
}\right]  =-n_{\mathbf{x}}\nabla
f\mathbf{\mathbf{\mathbf{_{\mathbf{\mathbf{\mathbf{\mathbf{\mathbf{\mathbf{x}%
}}}}}},}}} \label{cr}%
\end{equation}
valid for any c-number function
$f\mathbf{\mathbf{\mathbf{_{\mathbf{\mathbf{\mathbf{\mathbf{\mathbf{\mathbf{x}%
}}}}}}}}}$, we deduce
\begin{equation}
K_{1}=i\left[  Q,P\right]  =\int d\mathbf{\mathbf{x}}\widetilde{\mathbf{u}%
}_{\mathbf{x}}\int d\mathbf{y}Y\mathbf{_{\mathbf{y}}}i\left[
n\mathbf{_{\mathbf{y}},p}_{\mathbf{x}}\right]  =-\int d\mathbf{x}%
\widetilde{\mathbf{u}}_{\mathbf{x}}n_{\mathbf{x}}\nabla Y\mathbf{_{\mathbf{x}%
}.} \label{iQP1}%
\end{equation}
Using (\ref{rng}) yields after a partial integration
\begin{equation}
\left\langle K_{1}\right\rangle _{g}=-\int d\mathbf{x}\widetilde{\mathbf{u}%
}_{\mathbf{x}}\rho_{\mathbf{x}}\nabla Y\mathbf{_{\mathbf{x}}=}\int
d\mathbf{x}Y_{\mathbf{x}}\nabla\left(  \rho_{\mathbf{x}}%
\mathbf{\mathbf{\widetilde{\mathbf{u}}}_{\mathbf{\mathbf{x}}}}\right)  .
\label{K1g0}%
\end{equation}
The 'continuity equation' (\ref{qqe}) on accounting for (\ref{uw}) becomes
\begin{equation}
\partial_{q}\rho_{\mathbf{x}}=-\nabla\left(  \rho_{\mathbf{x}}\widetilde
{\mathbf{u}}_{\mathbf{\mathbf{x}}}\right)  , \label{ceq}%
\end{equation}
so that finally
\begin{equation}
\left\langle K_{1}\right\rangle _{g}=-\partial_{q}\int d\mathbf{x}%
Y_{\mathbf{x}}\rho_{\mathbf{x}}=-\partial_{q}q=-1. \label{K1g}%
\end{equation}
The second equality follows from (\ref{IYr=q}). From Eq. (\ref{iQP1}) and the
identity $n_{\mathbf{x}}=n_{\mathbf{x}}^{\prime}$ it follows that $K_{1}%
=K_{1}^{\prime}$, so that, invoking (\ref{trtr'}), (\ref{K1g}), we find the
equations $\left\langle K_{1}\right\rangle _{\mathrm{q}}=\left\langle
K_{1}\right\rangle _{g}=-1$, which owing to (\ref{iQP}) can be rewritten as
\begin{equation}
\left\langle \left[  Q,P\right]  \right\rangle _{\mathrm{q}}=\left\langle
\left[  Q,P\right]  \right\rangle _{g}=i. \label{cc}%
\end{equation}
These identities say that in a weak sense the operator $P$ can be regarded as
a momentum canonically conjugated to the coordinate $Q$.

Two last equalities in (\ref{trQPM}) lead to the relation $\left\langle
P\right\rangle _{\mathrm{q}}=\left\langle \mathcal{M}\right\rangle
_{\mathrm{q}}\dot{q}$ between the average values of $P$ and $\mathcal{M}$ at
fixed collective velocity $\dot{q}$ permitting one to anticipate that
$M=\left\langle \mathcal{M}\right\rangle _{\mathrm{q}}$ will represent the
microscopic or many-body expression for the collective inertia $B$.

While determining $\chi_{\mathbf{x}}$ and $\beta$ in the previous subsection
we postulated that expectation values $p_{n,t}=p_{n}(T_{i}+t)=$\textrm{tr}%
$\left(  P_{n,t}\rho_{\mathrm{q}}\right)  $ of Heisenberg representations of
collective operators $\left(  P_{1},P_{2},P_{3}\right)  =\left(
Q,P,\mathcal{M}\right)  $ vary with time much slower than $\rho_{\mathbf{x}%
,t}=$\textrm{tr}$\left(  n_{\mathbf{x},t}\rho_{\mathrm{q}}\right)  $ and
$\mathbf{j}_{\mathbf{x},t}=$\textrm{tr}$\left(  \mathbf{p}_{\mathbf{x},t}%
\rho_{\mathrm{q}}\right)  $. This postulate was argued by observation that
spatial averaging with proper weighting functions performed on $\rho
_{\mathbf{x},t}$ and $\mathbf{j}_{\mathbf{x},t}$ to get $p_{n,t}$, remove the
rapidly fluctuating components of $\rho_{\mathbf{x},t}$ and $\mathbf{j}%
_{\mathbf{x},t}$. As a result the time $t$ enters $p_{n,t}$ through
$q_{t}=q(T_{i}+t)$ and $\dot{q}_{t}=\partial q_{t}/\partial t$ only. Another
consequence of the hypothesis on slow character of collective motion,
expressed by Eq. (\ref{tq}), implies that $p_{n,t}$ are in fact the linear
functions of $t$ within the corresponding stage of collective motion. To
formulate one more implication of the slowness of $p_{n,t}$, to be employed in
finding them, it is convenient to put Eq. (\ref{frq}) for $\rho_{\mathrm{q}}$
into the form
\begin{equation}
\rho_{\mathrm{q}}=Z_{g}^{-1}e^{-\beta\left(  \mathcal{H}_{\mathrm{c}}-\mu
_{g}\mathcal{N}\right)  }, \label{rqS}%
\end{equation}
where%
\begin{equation}
\mathcal{H}_{\mathrm{c}}=\mathcal{H}+V, \label{Hcr}%
\end{equation}%
\begin{equation}
V=\Lambda_{g}Q-\dot{q}P+\frac{1}{2}\dot{q}^{2}\mathcal{M}. \label{V}%
\end{equation}
The rate of time change of $p_{n,t}$ is given by $\partial_{t}p_{n,t}%
=\mathrm{tr}\left(  \dot{P}_{n,t}\rho_{\mathrm{q}}\right)  $, where $\dot
{P}_{n,t}$ is the Heisenberg representation of the flux $\dot{P}_{n}\equiv
i\left[  \mathcal{H},P_{n}\right]  $. Using the cycling property of the trace,
the identity $i\left[  \mathcal{H}_{\mathrm{c}},\rho_{\mathrm{q}}\right]  =0$,
and Eq. (\ref{V}), we find
\[
\partial_{t}p_{n,t}=\mathrm{tr}\left(  i\left[  \mathcal{H},P_{n,t}\right]
\rho_{\mathrm{q}}\right)  =-\mathrm{tr}\left(  i\left[  \mathcal{H}%
,\rho_{\mathrm{q}}\right]  P_{n,t}\right)  =\mathrm{tr}\left(  i\left[
V,\rho_{\mathrm{q}}\right]  P_{n,t}\right)  .
\]
From this expression it follows that the necessary condition for the
adiabaticity of collective variables is that the quantity $i\left[
V,\rho_{\mathrm{q}}\right]  $ must be small, which does take place, as seen
from (\ref{rqS}), if $i\left[  V,\mathcal{H}_{\mathrm{c}}\right]  $ is small.

\section{The dynamic equation for $q\left(  T\right)  $}

\subsection{Collective fluxes}

In this section we intend to derive the dynamic equation for $q\left(
T\right)  $ from first principles. For our purposes it is necessary to know
the expectations of the fluxes $\dot{Q}$, $\dot{\mathcal{M}}$, $\dot{P}$ in
both $\rho_{g}$ and $\rho_{\mathrm{q}}$ ensembles. By definition
\[
\dot{Q}\equiv i\left[  \mathcal{H},Q\right]  ,\text{ \ \ \ \ }\dot
{\mathcal{M}}\equiv i\left[  \mathcal{H},\mathcal{M}\right]  ,\text{
\ \ \ \ }\dot{P}\equiv i\left[  \mathcal{H},P\right]  .
\]

Consider first the expectation value $\left\langle \dot{Q}\right\rangle
_{\mathrm{q}}$. Eq. (\ref{trtr'}) with $X=\dot{Q}$ in combination with the
relation $\dot{Q}=\dot{Q}^{\prime}-\dot{q}K_{1}^{\prime}$, proven in Appendix
A, (\ref{QdQd'}), enables us to deduce
\begin{equation}
\left\langle \dot{Q}\right\rangle _{\mathrm{q}}=\left\langle \dot
{Q}\right\rangle _{g}-\dot{q}\left\langle K_{1}\right\rangle _{g}.
\label{Qdq0}%
\end{equation}
In order to evaluate $\left\langle \dot{Q}\right\rangle _{g}$, we rewrite Eq.
(\ref{rhoeq}) in the form
\begin{equation}
\rho_{g}=Z_{g}^{-1}e^{-\beta(\mathcal{H}_{g}-\mu_{g}\mathcal{N})},
\label{rgHg}%
\end{equation}
where%
\begin{equation}
\mathcal{H}_{g}\equiv\mathcal{H}+\Lambda_{g}Q. \label{Hg}%
\end{equation}
Eq. (\ref{rgHg}) shows that $\mathcal{H}_{g}$ commutes with $\rho_{g}$. Using
this fact and the cyclic property of the trace, we find%
\begin{equation}
\left\langle \dot{Q}\right\rangle _{g}=\mathrm{tr}\left(  i\left[
\mathcal{H},Q\right]  \rho_{g}\right)  =\mathrm{tr}\left(  i\left[
\mathcal{H}_{g},Q\right]  \rho_{g}\right)  =\mathrm{tr}\left(  i\left[
\mathcal{H}_{g},Q\rho_{g}\right]  \right)  =0. \label{Qdg}%
\end{equation}
Substituting (\ref{Qdg}), (\ref{K1g}) into (\ref{Qdq0}), we finally obtain
\begin{equation}
\left\langle \dot{Q}\right\rangle _{\mathrm{q}}=\dot{q}. \label{Qdq}%
\end{equation}

The expectation value $\left\langle \dot{\mathcal{M}}\right\rangle
_{\mathrm{q}}$ of $\dot{\mathcal{M}}\equiv i\left[  \mathcal{H},\mathcal{M}%
\right]  $ is given by
\begin{equation}
\left\langle \dot{\mathcal{M}}\right\rangle _{\mathrm{q}}=\left\langle
\dot{\mathcal{M}}\right\rangle _{g}-\dot{q}\left\langle K_{2}\right\rangle
_{g}, \label{Mdq0}%
\end{equation}
where
\begin{equation}
K_{2}\equiv i\left[  \mathcal{M},P\right]  . \label{iMP}%
\end{equation}
To prove (\ref{Mdq0}) we make use of Eq. (\ref{trtr'}) at $X=\dot{\mathcal{M}%
}$ and employ the formula $\dot{\mathcal{M}}=\dot{\mathcal{M}}^{\prime}%
-\dot{q}K_{2}^{\prime}$, derived in Appendix A, (\ref{MdMd'}). From
(\ref{iMP}), (\ref{Mcoll}), (\ref{Pcoll}), and Eq. (\ref{cr}) with
$f\mathbf{\mathbf{\mathbf{_{\mathbf{\mathbf{\mathbf{\mathbf{\mathbf{\mathbf{x}%
}}}}}}=}}}$ $\widetilde{u}_{\mathbf{\mathbf{x}}}^{2}$ one finds
\begin{equation}
K_{2}=i\left[  \mathcal{M},P\right]  =m\int d\mathbf{x}d\mathbf{y}%
\widetilde{u}_{\mathbf{\mathbf{y}}}^{2}i\left[  n_{\mathbf{y}},\mathbf{p}%
_{\mathbf{x}}\right]  \mathbf{\widetilde{\mathbf{u}}}_{\mathbf{\mathbf{x}}%
}=-m\int d\mathbf{x}n_{\mathbf{x}}\mathbf{\widetilde{\mathbf{u}}%
}_{\mathbf{\mathbf{x}}}\nabla\widetilde{u}_{\mathbf{\mathbf{x}}}^{2},
\label{K2}%
\end{equation}
which on invoking (\ref{rng}) and integrating by parts with accounting for
(\ref{ceq}) yields
\begin{equation}
\left\langle K_{2}\right\rangle _{g}=-m\int d\mathbf{x}\widetilde
{u}_{\mathbf{\mathbf{x}}}^{2}\partial_{q}\rho_{\mathbf{x}}. \label{K2g}%
\end{equation}
Owing to the commutativity of $\mathcal{M}$ with $Q$, we can present
$\dot{\mathcal{M}}=i\left[  \mathcal{H},\mathcal{M}\right]  $ in the form
$\dot{\mathcal{M}}=i\left[  \mathcal{H}_{g},\mathcal{M}\right]  $. Using
$i\left[  \mathcal{H}_{g},\rho_{g}\right]  =0$ and the cyclic property of the
trace, we then find that%
\begin{equation}
\left\langle \dot{\mathcal{M}}\right\rangle _{g}=\mathrm{tr}\left(  i\left[
\mathcal{H}_{g},\mathcal{M}\right]  \rho_{g}\right)  =\mathrm{tr}\left(
i\left[  \mathcal{H}_{g},\mathcal{M}\rho_{g}\right]  \right)  =0. \label{Mdg}%
\end{equation}
Substituting Eqs. (\ref{Mdg}), (\ref{K2g}) into (\ref{Mdq0}), we finally
obtain
\begin{equation}
\left\langle \dot{\mathcal{M}}\right\rangle _{\mathrm{q}}=\dot{q}m\int
d\mathbf{x}\widetilde{u}_{\mathbf{\mathbf{x}}}^{2}\partial_{q}\rho
_{\mathbf{x}}. \label{Mdq}%
\end{equation}

The expectation value $\left\langle \dot{P}\right\rangle _{\mathrm{q}}$ of
$\dot{P}\equiv i\left[  \mathcal{H},P\right]  $ can be presented as
\begin{equation}
\left\langle \dot{P}\right\rangle _{\mathrm{q}}=\left\langle \dot
{P}\right\rangle _{g}+\dot{q}\left\langle \dot{\mathcal{M}}\right\rangle
_{g}-\frac{1}{2}\dot{q}^{2}\left\langle K_{2}\right\rangle _{g}=\left\langle
\dot{P}\right\rangle _{g}+\frac{1}{2}\dot{q}^{2}m\int d\mathbf{x}\widetilde
{u}_{\mathbf{\mathbf{x}}}^{2}\partial_{q}\rho_{\mathbf{x}}. \label{Pdq0}%
\end{equation}
The first equality is obtained using Eq. (\ref{trtr'}) with $X=\dot{P}$ and
Eq. (\ref{PdPd'}) from Appendix A, while the second one follows from
(\ref{Mdg}), (\ref{K2g}). Introducing the operator
\begin{equation}
\mathcal{F}\equiv i\left[  \mathcal{H}_{g},P\right]  , \label{Fc}%
\end{equation}
we utilize (\ref{Hg}), (\ref{iQP}) to present $\dot{P}\equiv i\left[
\mathcal{H},P\right]  $ in the form $\dot{P}=\mathcal{F}-\Lambda_{g}K_{1}$,
which on accounting for (\ref{K1g}) gives
\begin{equation}
\left\langle \dot{P}\right\rangle _{g}=\left\langle \mathcal{F}\right\rangle
_{g}+\Lambda_{g}. \label{Pdg}%
\end{equation}
Exploiting the identity $i\left[  \mathcal{H}_{g},\rho_{g}\right]  =0$ and the
cyclic property of the trace, we find%
\begin{equation}
\left\langle \mathcal{F}\right\rangle _{g}{}=\mathrm{tr}\left(  i\left[
\mathcal{H}_{g},P\right]  \rho_{g}\right)  =\mathrm{tr}\left(  i\left[
\mathcal{H}_{g},P\rho_{g}\right]  \right)  =0, \label{Fg}%
\end{equation}
which reduces Eq. (\ref{Pdg}) to
\begin{equation}
\left\langle \dot{P}\right\rangle _{g}=\Lambda_{g}, \label{Pdg1}%
\end{equation}
hence Eq. (\ref{Pdq0}) becomes
\begin{equation}
\left\langle \dot{P}\right\rangle _{\mathrm{q}}=\Lambda_{g}+\frac{1}{2}\dot
{q}^{2}m\int d\mathbf{x}\widetilde{u}_{\mathbf{\mathbf{x}}}^{2}\partial
_{q}\rho_{\mathbf{x}}. \label{Pdq}%
\end{equation}

According to (\ref{cc}) $P$ is canonically conjugated to $Q$ in a weak sense.
Therefore the $\dot{P}\equiv i\left[  \mathcal{H},P\right]  $ can be regarded
as operator of the deformation force. Thus, looking at Eq. (\ref{Pdg1}), which
indicates that the expectation of $\dot{P}$ with $\rho_{g}$ equals
$\Lambda_{g}$, we may anticipate that $\Lambda_{g}$ will represent the
microscopic expression for deformation force $f$.

\subsection{Microscopic or many-body expressions for $B$ and $f$}

Now we are prepared to derive the dynamic equation for $q(T)$. The derivation
is based not on the dynamic equation for $P_{t}=e^{it\mathcal{H}%
}Pe^{-it\mathcal{H}}$, as could be expected, but on that for operator
$K_{t}=e^{it\mathcal{H}}Ke^{-it\mathcal{H}}$, where
\begin{equation}
K=P-\dot{q}\mathcal{M}. \label{P2}%
\end{equation}
Since $\mathcal{M}$ commutes with $Q$, the $K$ operator is also canonically
conjugated to $Q$ in a weak sense, but it is easier than $P$ to deal with,
because, as follows from the relations $\mathrm{tr}\left(  P\rho_{\mathrm{q}%
}\right)  =M\dot{q}$,\ \ $\mathrm{tr}\left(  \mathcal{M}\rho_{\mathrm{q}%
}\right)  =M$, see (\ref{trQPM}), its expectation with $\rho_{\mathrm{q}}$
vanishes
\begin{equation}
\left\langle K\right\rangle _{\mathrm{q}}=\left\langle P\right\rangle
_{\mathrm{q}}-\dot{q}\left\langle \mathcal{M}\right\rangle _{\mathrm{q}}=0.
\label{trK}%
\end{equation}

The expectation of $K_{t}$ with $\rho_{\mathrm{q}}$ is designated as $k_{t}$:
\begin{equation}
k_{t}=\mathrm{tr}\left(  K_{t}\rho_{\mathrm{q}}\right)  . \label{pnt}%
\end{equation}
Differentiating (\ref{pnt}) with respect to $t$, we find
\begin{equation}
\partial_{t}k_{t}=\left\langle \dot{K}_{t}\right\rangle _{\mathrm{q}}.
\label{dpt}%
\end{equation}
Using (\ref{tq}) and (\ref{P2}) and remarking that $k_{0}=\mathrm{tr}\left(
K\rho_{\mathrm{q}}\right)  =0$ as seen from (\ref{trK}), we find that
$k_{t}=(t/\tau_{\mathrm{q}})k(\tau_{\mathrm{q}})$, which implies that
$\partial_{t}k_{t}=k(\tau_{\mathrm{q}})/\tau_{\mathrm{q}}$. Substituting this
expression into (\ref{dpt}) and integrating the resulting equation over $t$
with the weighting function $e^{-\eta t}$, in which $\eta=1/\tau_{\mathrm{q}}$
is the inverted duration time of the quasiequilibrium stage in question, we
obtain
\begin{equation}
k_{\tau_{\mathrm{q}}}=\int_{0}^{\infty}dte^{-\eta t}\left\langle \dot{K}%
_{t}\right\rangle _{\mathrm{q}}. \label{k1}%
\end{equation}
Eq. (\ref{k1}) provides a 'starting pad' for evaluating the dynamic equation
for $q(T)$ and obtaining the explicit many-body expressions for inertia $B$,
deformation force $f$, and friction $\gamma$.

In order to assess the left-hand side of (\ref{k1}), we substitute (\ref{P2})
into (\ref{pnt}) and use Eqs. (\ref{Pcoll}), (\ref{Mcoll}) for $P$,
$\mathcal{M}$ to find%
\begin{equation}
\text{\ }k_{t}=\int d\mathbf{x\mathbf{\widetilde{\mathbf{u}}}%
_{\mathbf{\mathbf{x}}}j}_{\mathbf{x},t}-\dot{q}m\int d\mathbf{x}\widetilde
{u}_{\mathbf{\mathbf{x}}}^{2}\rho_{\mathbf{x},t}, \label{p2t}%
\end{equation}
\ where $\rho_{\mathbf{x},t}\equiv\mathrm{tr}\left(  n_{\mathbf{x},t}%
\rho_{\mathrm{q}}\right)  $ and $\mathbf{j}_{\mathbf{x},t}\equiv
\mathrm{tr}\left(  \mathbf{p}_{\mathbf{x},t}\rho_{\mathrm{q}}\right)  $ are
defined in (\ref{rt}) and (\ref{jt}), respectively. In Eq. (\ref{p2t}) both
$\rho_{\mathbf{x},t}$ and $\mathbf{j}_{\mathbf{x},t}$ appear under the
integrals $\int d\mathbf{x...}$ over the entire nucleus. Therefore we may
replace them by their slow components in accordance with Eqs. (\ref{rxt}),
(\ref{uxt}). This gives
\begin{equation}
k_{\tau_{\mathrm{q}}}=\int d\mathbf{x\mathbf{\widetilde{\mathbf{u}}%
}_{\mathbf{\mathbf{x}}}}\rho_{A,\beta}(\mathbf{x},q_{\tau_{\mathrm{q}}}%
)m\dot{q}_{\tau_{\mathrm{q}}}\mathbf{\widetilde{\mathbf{u}}}_{A,\beta
}(\mathbf{x},q_{\tau_{\mathrm{q}}})-\dot{q}m\int d\mathbf{x}\widetilde
{u}_{\mathbf{\mathbf{x}}}^{2}\rho_{A,\beta}(\mathbf{x},q_{\tau_{\mathrm{q}}}).
\label{k}%
\end{equation}
If we introduce the variations
\[
\delta\rho_{\mathbf{x}}\equiv\rho_{A,\beta}(\mathbf{x},q_{\tau_{\mathrm{q}}%
})-\rho_{\mathbf{x}},\text{ \ \ \ \ }\delta\dot{q}\equiv\dot{q}_{\tau
_{\mathrm{q}}}-\dot{q},\text{ \ \ \ \ \ }\delta\mathbf{\widetilde{\mathbf{u}}%
}_{\mathbf{\mathbf{x}}}\equiv\mathbf{\widetilde{\mathbf{u}}}_{A,\beta
}(\mathbf{x},q_{\tau_{\mathrm{q}}})-\mathbf{\widetilde{\mathbf{u}}%
}_{\mathbf{\mathbf{x}}},
\]
then, in the lowest order in those variations, we obtain%
\[
k_{\tau_{\mathrm{q}}}=\dot{q}m\int d\mathbf{x}\rho_{\mathbf{x}}%
\mathbf{\mathbf{\widetilde{\mathbf{u}}}_{\mathbf{\mathbf{x}}}}\delta
\mathbf{\widetilde{\mathbf{u}}}_{\mathbf{\mathbf{x}}}+m\int d\mathbf{x}%
\rho_{\mathbf{x}}\widetilde{u}_{\mathbf{\mathbf{x}}}^{2}\delta\dot{q}.
\]
Approximating $\delta\mathbf{\widetilde{\mathbf{u}}}_{\mathbf{\mathbf{x}}}$ by
$\left(  \partial_{q}\mathbf{\widetilde{\mathbf{u}}}_{\mathbf{\mathbf{x}}%
}\right)  \delta q$, where\ $\delta q\equiv q_{\tau_{\mathrm{q}}}-q$, and
using Eq. (\ref{Mq}) for $M$, we arrive at the desired estimate of the left
side of (\ref{k1}):%
\begin{equation}
k_{\tau_{\mathrm{q}}}=\frac{1}{2}\dot{q}m\int d\mathbf{x}\rho_{\mathbf{x}%
}\left(  \partial_{q}\widetilde{u}_{\mathbf{\mathbf{x}}}^{2}\right)  \delta
q+M\delta\dot{q}. \label{ktq}%
\end{equation}

The following evaluation of the right side of (\ref{k1}) is based on a
formally exact integral equation for $\dot{K}_{t}$:%
\begin{equation}
\dot{K}_{t}=e^{it\mathcal{H}_{\mathrm{c}}}\dot{K}e^{-it\mathcal{H}%
_{\mathrm{c}}}+\int_{0}^{t}dse^{i(t-s)\mathcal{H}_{\mathrm{c}}}i\left[
\dot{K}_{s},V\right]  e^{-i(t-s)\mathcal{H}_{\mathrm{c}}}, \label{AP4}%
\end{equation}
which is derived in Appendix B (see also \cite{AP}). By making use of the
cycling property of the trace and taking into account that $\rho_{\mathrm{q}}$
commutes with $\mathcal{H}_{\mathrm{c}}$, we prove for any operator $X$ the
relationship
\begin{equation}
\mathrm{tr}\left(  e^{it\mathcal{H}_{\mathrm{c}}}Xe^{-it\mathcal{H}%
_{\mathrm{c}}}\rho_{\mathrm{q}}\right)  =\mathrm{tr}\left(  Xe^{-it\mathcal{H}%
_{\mathrm{c}}}\rho_{\mathrm{q}}e^{it\mathcal{H}_{\mathrm{c}}}\right)
=\mathrm{tr}\left(  X\rho_{\mathrm{q}}\right)  . \label{rs}%
\end{equation}
Taking the expectations of both sides of (\ref{AP4}) and using (\ref{rs}) we
find that%
\begin{equation}
\left\langle \dot{K}_{t}\right\rangle _{\mathrm{q}}=\left\langle \dot
{K}\right\rangle _{\mathrm{q}}+\int_{0}^{t}ds\left\langle i\left[  \dot{K}%
_{s},V\right]  \right\rangle _{\mathrm{q}}. \label{Kdtq}%
\end{equation}
From Eqs. (\ref{P2}), (\ref{Pdq}), (\ref{Mdq}) we obtain
\begin{equation}
\left\langle \dot{K}\right\rangle _{\mathrm{q}}=\Lambda_{g}-\frac{1}{2}\dot
{q}^{2}m\int d\mathbf{x}\widetilde{u}_{\mathbf{\mathbf{x}}}^{2}\partial
_{q}\rho_{\mathbf{x}}, \label{Kdq3}%
\end{equation}
so that (\ref{Kdtq}) gets transformed into
\begin{equation}
\left\langle \dot{K}_{t}\right\rangle _{\mathrm{q}}=\Lambda_{g}-\frac{1}%
{2}\dot{q}^{2}m\int d\mathbf{x}\widetilde{u}_{\mathbf{\mathbf{x}}}^{2}%
\partial_{q}\rho_{\mathbf{x}}+\int_{0}^{t}ds\left\langle i\left[  \dot{K}%
_{s},V\right]  \right\rangle _{\mathrm{q}}. \label{Kdtq1}%
\end{equation}

Substituting Eqs. (\ref{ktq}), (\ref{Kdtq1}) for $k_{\tau_{\mathrm{q}}}$,
$\left\langle \dot{K}_{t}\right\rangle _{\mathrm{q}}$, respectively, into
(\ref{k1}), multiplying both sides of the resulting equation by $\eta
=\tau_{\mathrm{q}}^{-1}$ and introducing the notations
\[
q^{\prime\prime}=\delta\dot{q}/\tau_{\mathrm{q}}=\left(  \dot{q}%
_{\tau_{\mathrm{q}}}-\dot{q}\right)  /\tau_{\mathrm{q}},\ \text{\ \ \ \ }%
q^{\prime}=\delta q/\tau_{\mathrm{q}}=\left(  q_{\tau_{\mathrm{q}}}-q\right)
/\tau_{\mathrm{q}},
\]
we obtain%
\begin{equation}
\frac{1}{2}\dot{q}m\int d\mathbf{x}\rho_{\mathbf{x}}\left(  \partial
_{q}\widetilde{u}_{\mathbf{\mathbf{x}}}^{2}\right)  q^{\prime}+Mq^{\prime
\prime}=\Lambda_{g}-\frac{1}{2}\dot{q}^{2}m\int d\mathbf{x}\widetilde
{u}_{\mathbf{\mathbf{x}}}^{2}\partial_{q}\rho_{\mathbf{x}}+r(\tau_{\mathrm{q}%
}), \label{deq0}%
\end{equation}
with
\begin{equation}
r(\tau_{\mathrm{q}})=\eta\int_{0}^{\infty}dte^{-\eta t}\int_{0}^{t}%
ds\left\langle i\left[  \dot{K}_{s},V\right]  \right\rangle _{\mathrm{q}}%
=\int_{0}^{\infty}dte^{-\eta t}\left\langle i\left[  \dot{K}_{t},V\right]
\right\rangle _{\mathrm{q}}. \label{rtq2}%
\end{equation}
The second expression in (\ref{rtq2}) has been derived from the first one by
using the identity $\eta e^{-\eta t}=-\partial_{t}e^{-\eta t}$ and integrating
by parts. After using the relation
\[
\text{\ }\partial_{q}M=m\int d\mathbf{x}\left(  \rho_{\mathbf{x}}\partial
_{q}\widetilde{u}_{\mathbf{\mathbf{x}}}^{2}+\widetilde{u}_{\mathbf{\mathbf{x}%
}}^{2}\partial_{q}\rho_{\mathbf{x}}\right)  ,
\]
which follows from (\ref{Mq}), and replacing $q^{\prime}$ with $\dot{q}$ and
$q^{\prime\prime}$ with $\ddot{q}$, Eq. (\ref{deq0}) reduces to
\begin{equation}
\ddot{q}M+\frac{1}{2}\dot{q}{}^{2}\partial_{q}M=\Lambda_{g}+r(\tau
_{\mathrm{q}}), \label{deq}%
\end{equation}
where $M$ and $\Lambda_{g}$ are defined in (\ref{Mq}) and (\ref{mL}), respectively.

The composition of Eq. (\ref{deq}) would be fully identical to that of the
dynamic equation (\ref{deI}) for $q(T)$, if $r(\tau_{\mathrm{q}})$ in
(\ref{deq}) were linear in $\dot{q}$. In subsection 3.3 we show that in the
adiabatic approximation this is the case. Therefore we find the following
many-body expressions for inertia $B$ and deformation force $f$ in
(\ref{deI}):
\begin{equation}
B=M,\text{ \ \ \ \ \ \ }f=\Lambda_{g}, \label{Bf}%
\end{equation}
which confirm our expectations concerning $M$ and $\Lambda_{g}$.

In order to find an alternative expression for $f$, we prove the formula
\begin{equation}
\Lambda_{g}=-\partial_{q}F, \label{Lamb2}%
\end{equation}
where the $F$ is Helmholtz's free energy:
\begin{equation}
F=U-\beta^{-1}S. \label{Fe}%
\end{equation}
As a first step we use Eq. (\ref{rhoeq}), the normalization condition
$\mathrm{tr}\rho_{g}=1$, Eq. (\ref{U1}), and equations (\ref{NQ}) at $\mu
=\mu_{g}$, $\Lambda=\Lambda_{g}$, to put the entropy $S=-\mathrm{tr}(\rho
_{g}\ln\rho_{g})$ into the form
\begin{equation}
S=\ln Z_{g}+\beta U-\beta\mu_{g}A+\beta\Lambda_{g}q. \label{Sg1}%
\end{equation}
Next we put this into (\ref{Fe}) and differentiate the result over $q$, which
yields
\begin{equation}
-\partial_{q}F=\Lambda_{g}+\beta^{-1}\partial_{q}\ln Z_{g}+q\partial
_{q}\Lambda_{g}-A\partial_{q}\mu_{g}. \label{dqF}%
\end{equation}
From Eq. (\ref{Zg}) and equations in (\ref{NQ}) with $\mu=\mu_{g}$,
$\Lambda=\Lambda_{g}$ we derive the relation
\[
\beta^{-1}\partial_{q}\ln Z_{g}=\beta^{-1}Z_{g}^{-1}\mathrm{tr}\partial
_{q}e^{-\beta\left(  \mathcal{H}+\Lambda_{g}Q-\mu_{g}\mathcal{N}\right)
}=-\mathrm{tr}\left[  \rho_{g}\partial_{q}\left(  \Lambda_{g}Q-\mu
_{g}\mathcal{N}\right)  \right]  =-q\partial_{q}\Lambda_{g}+A\partial_{q}%
\mu_{g},
\]
which reduces (\ref{dqF}) to (\ref{Lamb2}).

According to the expressions
\[
f=-\partial_{q}W,\text{ \ \ \ }f=\Lambda_{g},\text{ \ \ \ }\Lambda
_{g}=-\partial_{q}F,
\]
the deformation potential $W$ must be identified (up to an additive constant)
with Helmholtz's free energy $F$. This justifies the practical procedures of
calculating $W$ in hot nuclei described in \cite{RKNA},\ and the references
cited therein.

\subsection{Many-body expressions for $\gamma$}

In order to facilitate the analysis of $r(\tau_{\mathrm{q}})$, the flux
$\dot{K}_{t}$ in (\ref{rtq2}) is rewritten in the form
\begin{equation}
\dot{K}_{t}=\dot{K}_{t}^{\mathrm{c}}-\Lambda_{g}K_{1,t}+\frac{1}{2}\dot{q}%
^{2}K_{2,t}, \label{Ktd}%
\end{equation}
with
\begin{equation}
\dot{K}^{\mathrm{c}}\equiv i\left[  \mathcal{H}_{\mathrm{c}},K\right]  .
\label{Pndc}%
\end{equation}
Eq. (\ref{Ktd}) follows from Eqs. (\ref{Hcr}), (\ref{P2}), $\left[
Q,\mathcal{M}\right]  =0$, (\ref{iQP}), (\ref{iMP}). Since $\rho_{\mathrm{q}}$
commutes with $\mathcal{H}_{\mathrm{c}}$, we have $(n=1,2)$%
\begin{equation}
\mathrm{tr}\left(  i\left[  \mathcal{H}_{\mathrm{c}},K_{n,t}\right]
\rho_{\mathrm{q}}\right)  =\mathrm{tr}\left(  i\left[  \mathcal{H}%
_{\mathrm{c}},K_{n,t}\rho_{\mathrm{q}}\right]  \right)  =0. \label{cp}%
\end{equation}
The last equality follows from the general relation $\mathrm{tr}\left(
[X,Y]\right)  =0$ due to the cyclic property of the trace. Combining
(\ref{Hcr}) with (\ref{cp}) allows to write:%
\begin{equation}
\mathrm{tr}\left(  i\left[  K_{n,t},V\right]  \rho_{\mathrm{q}}\right)
=\mathrm{tr}\left(  i\left[  K_{n,t},\mathcal{H}_{\mathrm{c}}-\mathcal{H}%
\right]  \rho_{\mathrm{q}}\right)  =\mathrm{tr}\left(  \dot{K}_{n,t}%
\rho_{\mathrm{q}}\right)  , \label{VH}%
\end{equation}
where $\dot{K}_{n,t}=i\left[  \mathcal{H},K_{n,t}\right]  $. Substituting
(\ref{Ktd}) into (\ref{rtq2}) and accounting for (\ref{VH}), we obtain%
\begin{equation}
r(\tau_{\mathrm{q}})=\int_{0}^{\infty}dte^{-\eta t}\left\langle i\left[
\dot{K}_{t}^{\mathrm{c}}{},V\right]  \right\rangle _{\mathrm{q}}-\Lambda
_{g}\int_{0}^{\infty}dte^{-\eta t}\left\langle {}\dot{K}_{1,t}\right\rangle
_{\mathrm{q}}+\frac{1}{2}\dot{q}^{2}\int_{0}^{\infty}dte^{-\eta t}\left\langle
{}\dot{K}_{2,t}\right\rangle _{\mathrm{q}}. \label{rtq3}%
\end{equation}

According to Appendix C, the integrals $\int_{0}^{\infty}dte^{-\eta
t}\left\langle {}\dot{K}_{n,t}\right\rangle _{\mathrm{q}}$, $n=1,2$, are
proportional to $\dot{q}$, which is a small quantity of our problem. Therefore
the second and third terms in (\ref{rtq3}) lead to the corrections to
$\Lambda_{g}$ and $\frac{1}{2}\dot{q}{}^{2}\partial_{q}M$ in (\ref{deq}),
which can be neglected. With this in mind we find%
\[
r(\tau_{\mathrm{q}})=\int_{0}^{\infty}dte^{-\eta t}\mathrm{tr}\left(  i\left[
\dot{K}_{t}^{\mathrm{c}}{},V\right]  \rho_{\mathrm{q}}{}\right)  ,
\]
or, using $\dot{K}_{t}^{\mathrm{c}}{}=e^{it\mathcal{H}}\dot{K}^{\mathrm{c}}%
{}e^{-it\mathcal{H}}$ and the cyclic property of the trace
\begin{equation}
r(\tau_{\mathrm{q}})=-\int_{0}^{\infty}dte^{-\eta t}\mathrm{tr}\left(
e^{it\mathcal{H}}\dot{K}^{\mathrm{c}}{}e^{-it\mathcal{H}}i\left[
\rho_{\mathrm{q}}{},V\right]  \right)  . \label{rtq6}%
\end{equation}
Regarding $V$ as a small perturbation of $\mathcal{H}_{\mathrm{c}}$ in the
expression $\mathcal{H}=\mathcal{H}_{\mathrm{c}}-V$ enables one to replace the
operators $e^{\pm it\mathcal{H}}=$ $e^{\pm it\left(  \mathcal{H}_{\mathrm{c}%
}-V\right)  }$ in (\ref{rtq6}) with $e^{\pm it\mathcal{H}_{\mathrm{c}}}$,
because an error introduced into $r(\tau_{\mathrm{q}})$ by this replacement is
of second order in $V{}$. This gives%
\begin{equation}
r(\tau_{\mathrm{q}})=-\int_{0}^{\infty}dte^{-\eta t}\mathrm{tr}\left(
e^{it\mathcal{H}_{\mathrm{c}}}\dot{K}^{\mathrm{c}}{}e^{-it\mathcal{H}%
_{\mathrm{c}}}i\left[  \rho_{\mathrm{q}}{},V\right]  \right)  =\int
_{0}^{\infty}dte^{-\eta t}\left\langle i\left[  \dot{K}^{\mathrm{c}%
},e^{-it\mathcal{H}_{\mathrm{c}}}{}Ve^{it\mathcal{H}_{\mathrm{c}}}\right]
\right\rangle _{\mathrm{q}}. \label{rq}%
\end{equation}
The second expression is obtained from the first by cycling permutation of
operators inside the trace. According to Eqs. (\ref{Hcr4}), (\ref{V'}),
(\ref{Kdc}) of Appendix A, the $\mathcal{H}_{\mathrm{c}}$, $V$, $\dot
{K}^{\mathrm{c}}$ are related to the primed operators as follows%
\begin{equation}
\mathcal{H}_{\mathrm{c}}=\mathcal{H}_{g}^{\prime},\text{ \ \ }V=\Lambda
_{g}Q^{\prime}-\dot{q}P^{\prime}-\frac{1}{2}\dot{q}^{2}\mathcal{M}^{\prime
},\text{ \ \ }\dot{K}^{\mathrm{c}}=\mathcal{F}^{\prime}. \label{rq3}%
\end{equation}
Inserting these expressions into (\ref{rq}), accounting for (\ref{trtr'}), and
removing all primes, we find%
\begin{equation}
r(\tau_{\mathrm{q}})=\int_{0}^{\infty}dte^{-\eta t}\left\langle i\left[
\mathcal{F},e^{-it\mathcal{H}_{g}}{}\left(  \Lambda_{g}Q-\dot{q}P-\frac{1}%
{2}\dot{q}^{2}\mathcal{M}\right)  e^{it\mathcal{H}_{g}}\right]  \right\rangle
_{g}. \label{rg}%
\end{equation}

According to (\ref{rg}) $r(\tau_{\mathrm{q}})$ contains in addition to the
term proportional to $\dot{q}$ also the terms proportional to $\Lambda_{g}$
and $\dot{q}^{2}$. Evidently, these additional terms, upon putting (\ref{rg})
into (\ref{deq}), will modify $\Lambda_{g}$ and $\frac{1}{2}\dot{q}{}%
^{2}\partial_{q}M$. Since these modifications would destroy our expressions
(\ref{Bf}) for $B$ and $f$, whose validity is beyond doubt, we require that
the coefficients in front of the terms proportional to $\Lambda_{g}$ and
$\dot{q}^{2}$ are negligible small:
\begin{equation}
\left\vert \int_{0}^{\infty}dte^{-\eta t}\left\langle \left[  \mathcal{F}%
,e^{-it\mathcal{H}_{g}}{}Qe^{it\mathcal{H}_{g}}\right]  \right\rangle
_{g}\right\vert \ll1,\ \ \ \ \left\vert \int_{0}^{\infty}dte^{-\eta
t}\left\langle \left[  \mathcal{F},e^{-it\mathcal{H}_{g}}{}\mathcal{M}%
e^{it\mathcal{H}_{g}}\right]  \right\rangle _{g}\right\vert \ll\left\vert
\partial_{q}M\right\vert . \label{<<}%
\end{equation}
As a result, we obtain the dynamic equation for $q(T)$ in the final form:%
\begin{equation}
\ddot{q}M+\frac{1}{2}\dot{q}{}^{2}\partial_{q}M=\Lambda_{g}-\dot{q}\gamma,
\label{de}%
\end{equation}
where%
\begin{equation}
\gamma=\int_{0}^{\infty}dte^{-\eta t}\left\langle i\left[  \mathcal{F}%
,e^{-it\mathcal{H}_{g}}{}Pe^{it\mathcal{H}_{g}}\right]  \right\rangle _{g}.
\label{g00}%
\end{equation}

The physical condition for the validity of (\ref{<<}) is the adiabaticity of
collective motion. To show this we note that adiabaticity implies that $Q$ and
$\mathcal{M}$ are 'quasiintegrals of motion', which means that their
commutators with $\mathcal{H}$, which are equal to those with $\mathcal{H}%
_{g}$, are small. Letting $\left[  \mathcal{H}_{g},Q\right]  =\left[
\mathcal{H}_{g},\mathcal{M}\right]  =0$, we find that $e^{-it\mathcal{H}_{g}%
}{}Qe^{it\mathcal{H}_{g}}=Q$, $e^{-it\mathcal{H}_{g}}{}\mathcal{M}%
e^{it\mathcal{H}_{g}}=\mathcal{M}$ in the static limit. The identities
$\left\langle \left[  \mathcal{F}{},Q\right]  \right\rangle _{g}=0$,
$\left\langle \left[  \mathcal{F}{},\mathcal{M}\right]  \right\rangle _{g}=0$,
proven in Appendix D, then show that in this limit%
\[
\int_{0}^{\infty}dte^{-\eta t}\left\langle \left[  \mathcal{F}%
,e^{-it\mathcal{H}_{g}}{}Qe^{it\mathcal{H}_{g}}\right]  \right\rangle
_{g}=0,\text{ \ \ \ \ \ \ \ }\int_{0}^{\infty}dte^{-\eta t}\left\langle
\left[  \mathcal{F},e^{-it\mathcal{H}_{g}}{}\mathcal{M}e^{it\mathcal{H}_{g}%
}\right]  \right\rangle _{g}=0.
\]
These identities confirm that for sufficiently slow collective motion the
conditions (\ref{<<}) are satisfied.

Inserting into (\ref{g00}) the identity $1=\partial_{t}t$, integrating by
parts and using the relations%
\[
\partial_{t}e^{-\eta t}=-\eta e^{-\eta t},\text{ \ \ \ \ \ \ \ }\partial
_{t}e^{-it\mathcal{H}_{g}}Pe^{it\mathcal{H}_{g}}=-e^{-it\mathcal{H}_{g}%
}i\left[  \mathcal{H}_{g},P\right]  e^{it\mathcal{H}_{g}}=-e^{-it\mathcal{H}%
_{g}}\mathcal{F}e^{it\mathcal{H}_{g}},
\]
the last of which follows from (\ref{Fc}), we obtain%
\begin{equation}
\gamma=\eta\int_{0}^{\infty}dte^{-\eta t}t\left\langle i\left[  \mathcal{F}%
,e^{-it\mathcal{H}_{g}}{}Pe^{it\mathcal{H}_{g}}\right]  \right\rangle
_{g}+\int_{0}^{\infty}dtte^{-\eta t}\left\langle i\left[  \mathcal{F}%
,e^{-it\mathcal{H}_{g}}\mathcal{F}e^{it\mathcal{H}_{g}}\right]  \right\rangle
_{g}. \label{g01}%
\end{equation}
In the classical case one describes the nucleus in terms of the coordinates
$\mathbf{x}_{j}$ and momenta $\mathbf{p}_{j}$ of the particles $j=1,...A$,
bound by the mean field potential $V(\mathbf{x},q)$. The $P$ and
$\mathcal{F}=i\left[  \mathcal{H}_{g},P\right]  $ are given in this case by
$\sum\widetilde{p}_{j}$ and $\sum\widetilde{F}_{j}$, respectively, with
$\widetilde{p}_{j}=\widetilde{\mathbf{u}}_{\mathbf{x}_{j}}\mathbf{p}_{j}%
$,\ $\widetilde{F}_{j}=-\widetilde{\mathbf{u}}_{\mathbf{x}_{j}}\nabla
V(\mathbf{x}_{j},q)$. Recalling the definition $\eta=1/\tau_{\mathrm{q}}$, we
see then that the first term in (\ref{g01}) is smaller than the second one by
a factor of $T/\tau_{\mathrm{q}}$, with $T\sim\sum\widetilde{p}_{j}%
/\sum\widetilde{F}_{j}$ denoting the characteristic period of particle motion
in the direction of the mass flow. As far as characteristic times of the
internal motion are much shorter than $\tau_{\mathrm{q}}$, we may omit the
first term in (\ref{g01}) to obtain%
\[
\gamma=i\int_{0}^{\infty}dte^{-\eta t}t\mathrm{tr}\left(  \mathcal{F}%
e^{-it\mathcal{H}_{g}}\mathcal{F}e^{it\mathcal{H}_{g}}\rho_{g}\right)
-i\int_{0}^{\infty}dte^{-\eta t}t\mathrm{tr}\left(  e^{-it\mathcal{H}_{g}%
}\mathcal{F}e^{it\mathcal{H}_{g}}\mathcal{F}\rho_{g}\right)  .
\]
Now we use in the second term the cyclic property of the trace, the
commutativity of $\mathcal{H}_{g}$ with $\rho_{g}$ and the substitution
$t\rightarrow-t$ to join it to the first one:%
\begin{equation}
\gamma=i\int_{-\infty}^{\infty}dte^{-\eta|t|}tC_{\mathcal{FF}}\left(
t\right)  , \label{gc}%
\end{equation}
where $C_{\mathcal{FF}}\left(  t\right)  $ is the time correlation function of
$\mathcal{F}$:%
\begin{equation}
C_{\mathcal{FF}}\left(  t\right)  =\mathrm{tr}\left(  e^{it\mathcal{H}_{g}%
}\mathcal{F}e^{-it\mathcal{H}_{g}}\mathcal{F}\rho_{g}\right)  =\mathrm{tr}%
\left(  e^{it\mathcal{H}_{g}}\mathcal{F}e^{-it\mathcal{H}_{g}}\mathcal{F}%
\rho_{g}\right)  . \label{CFF}%
\end{equation}

Note that the Hamiltonian $\mathcal{H}_{g}$, generating the time evolution of
the first $\mathcal{F}$ in (\ref{CFF}), differs from the true Hamiltonian
$\mathcal{H}$ by a term $\Lambda_{g}Q$, which is the analog of the
electron-nuclei interaction component of electronic Hamiltonian in the diatom
visualization of nuclear collective motion. The Hamiltonian $\mathcal{H}_{g}$
enters $C_{\mathcal{FF}}\left(  t\right)  $ not only through $\rho_{g}$ and
$e^{\pm it\mathcal{H}_{g}}$, but also through $\mathcal{F}=i\left[
\mathcal{H}_{g},P\right]  $, in which the $P$ itself can be put into the form
$P=mi[\mathcal{H}_{g},\widetilde{\phi}]$, where $\widetilde{\phi}\equiv\int
d\mathbf{x}\widetilde{\phi}\mathbf{_{\mathbf{x}}}n_{\mathbf{x}}$. This
expression for $P$ is proven by using Eq. (\ref{pot}) for $\mathbf{u}%
_{\mathbf{x}}$ in Eq. (\ref{Pcoll}) for $P$, integrating by parts, and
employing the relation $\nabla\mathbf{p}_{\mathbf{x}}=-mi[\mathcal{H}%
_{g},n_{\mathbf{x}}]$, which follows from Eq. (\ref{nd}), because of
commutativity of $Q$ with $n_{\mathbf{x}}$. Using the above expressions for
$\mathcal{F}$ and $P$, we put $C_{\mathcal{FF}}\left(  t\right)  $ into the
form
\begin{equation}
C_{\mathcal{FF}}\left(  t\right)  =m^{2}\frac{d^{4}C_{\widetilde{\phi
}\widetilde{\phi}}\left(  t\right)  }{dt^{4}},\text{ \ \ \ \ \ \ \ }%
C_{\widetilde{\phi}\widetilde{\phi}}\left(  t\right)  =\mathrm{tr}\left(
e^{it\mathcal{H}_{g}}\widetilde{\phi}e^{-it\mathcal{H}_{g}}\widetilde{\phi
}\rho_{g}\right)  . \label{Cff}%
\end{equation}
Combining Eqs. (\ref{gc}), (\ref{Cff}), we obtain the expression
\begin{equation}
\gamma=im^{2}\int_{-\infty}^{\infty}dte^{-\eta|t|}tC_{\widetilde{\phi
}\widetilde{\phi}}^{\prime\prime\prime\prime}\left(  t\right)  ,\text{
\ \ \ \ \ \ }C_{\widetilde{\phi}\widetilde{\phi}}^{\prime\prime\prime\prime
}\left(  t\right)  =\frac{d^{4}C_{\widetilde{\phi}\widetilde{\phi}}\left(
t\right)  }{dt^{4}}, \label{gf}%
\end{equation}
relating friction with the time correlation function of$\ $the reduced phase
operator $\widetilde{\phi}=\left(  \dot{q}m\right)  ^{-1}\chi$ in the
$\rho_{g}$ ensemble.

\section{Collective dynamics of self-confined Fermi gas}

\subsection{ Deformation force}

For practical calculations of key ingredients ($f$, $B$, $\gamma$) of equation
(\ref{deI}) we treat the nucleus as a self-confined Fermi gas of
quasiparticles bound by a self-consistent potential generated by the effective
force. The quasiparticle energies are denoted as $E_{\lambda}$, their
real-valued wave functions being $\varphi_{\lambda}\left(  \mathbf{x}\right)
$. According to \cite{BalianDominicis,UMT}, the entropy $S$ and the one-body
density matrix $\rho_{\mathbf{yz}}=\left\langle \psi_{\mathbf{y}}^{\ast}%
\psi_{\mathbf{z}}\right\rangle $ in this gas have the same expressions in
terms of occupation numbers $f_{\lambda}$ as for noninteracting fermions
\begin{equation}
S=-\sum_{\lambda}\left[  f_{\lambda}\ln f_{\lambda}+\left(  1-f_{\lambda
}\right)  \ln\left(  1-f_{\lambda}\right)  \right]  , \label{S}%
\end{equation}%
\begin{equation}
\rho_{\mathbf{yz}}=\sum_{\lambda}f_{\lambda}\varphi_{\lambda}\left(
\mathbf{y}\right)  \varphi_{\lambda}\left(  \mathbf{z}\right)  , \label{rxy}%
\end{equation}
where
\begin{equation}
f_{\lambda}=\left(  1+e^{\beta\left(  E_{\lambda}-\mu\right)  }\right)  ^{-1}.
\label{fl}%
\end{equation}
The Gibbs grand potential $\Omega\left[  \rho\right]  $ of the system with
Hamiltonian $\mathcal{H}_{g}$ and density matrix $\rho$, is given by%
\begin{equation}
\Omega\left[  \rho\right]  =\mathrm{tr}\left[  \left(  \mathcal{H}_{g}%
-\mu\mathcal{N}\right)  \rho\right]  -\beta^{-1}S\left[  \rho\right]  ,
\label{G}%
\end{equation}
where $S\left[  \rho\right]  =-\mathrm{tr}(\rho\ln\rho)$ is the entropy of the
$\rho$ ensemble. A remarkable property of $\Omega\left[  \rho\right]  $,
established by Gibbs in the classical case, is that it reaches its minimal
value $\Omega_{\min}$ for $\rho=\rho_{g}$. For quantum case this property is
proven in \cite{Neumann} and \cite{Mermin}. It follows from (\ref{Sg1}),
(\ref{Hg}), (\ref{U1}), (\ref{NQ}) that
\[
\Omega_{\min}=\Omega\left[  \rho_{g}\right]  =-\beta^{-1}\ln Z_{g}.
\]

Let $\rho$ differs from $\rho_{g}$ by small variations of wave functions
$\varphi_{\lambda}\left(  \mathbf{x}\right)  $, but has the same values of
$\beta$, $\mu$, $\Lambda$, $E_{\lambda}$ as in $\rho_{g}$. Since $\beta
^{-1}S\left[  \rho\right]  $ for so defined $\rho$ coincides with $\beta
^{-1}S$, because $S$ according to (\ref{S}), (\ref{fl}) is not depending on
$\varphi_{\lambda}\left(  \mathbf{x}\right)  $, the variational derivative of
$\Omega\left[  \rho\right]  $ over $\varphi_{\nu}\left(  \mathbf{x}\right)  $
coincides with that of $\mathrm{tr}\left[  \left(  \mathcal{H}_{g}%
-\mu\mathcal{N}\right)  \rho\right]  $. Using Eqs. (\ref{Hg}) and (\ref{Nc})
for $\mathcal{H}_{g}$ and $\mathcal{N}$, respectively, the variational
equations for $\varphi_{\nu}\left(  \mathbf{x}\right)  $ can consequently be
written as
\begin{equation}
\frac{\delta}{\delta\varphi_{\nu}\left(  \mathbf{x}\right)  }\left(  U\left[
\rho\right]  +\Lambda\int d\mathbf{y}Y_{\mathbf{y}}\rho_{\mathbf{y}}-\mu\int
d\mathbf{y}\rho_{\mathbf{y}}-\sum_{\lambda}f_{\lambda}e_{\lambda}\int
d\mathbf{y}\varphi_{\lambda}^{2}\left(  \mathbf{y}\right)  \right)  =0,
\label{d(U-)}%
\end{equation}
where $U\left[  \rho\right]  =\mathrm{tr}\left(  \mathcal{H}\rho\right)  $
and
\begin{equation}
\rho_{\mathbf{y}}=\sum_{\lambda}f_{\lambda}\varphi_{\lambda}^{2}\left(
\mathbf{y}\right)  \label{ryHF}%
\end{equation}
are the internal energy and the number density in the $\rho$ ensemble,
respectively, while $f_{\lambda}e_{\lambda}$ are the Lagrange parameters to
ensure that each wave function $\varphi_{\lambda}\left(  \mathbf{y}\right)  $
should remain normalized under variations.

In order to present $U\left[  \rho\right]  =\mathrm{tr}\left(  \mathcal{H}%
\rho\right)  $ as a functional of $\varphi_{\lambda}$, we first use Eq.
(\ref{Hcal}) for $\mathcal{H}$ to write
\begin{equation}
U\left[  \rho\right]  =-\frac{1}{2m}\int d\mathbf{y}\left\langle
\psi_{\mathbf{y}}^{\ast}\Delta_{\mathbf{y}}\psi_{\mathbf{y}}\right\rangle
+\frac{1}{2}\int d\mathbf{y}d\mathbf{z}v(|\mathbf{y}-\mathbf{z}|)\left\langle
\psi_{\mathbf{y}}^{\ast}\psi_{\mathbf{z}}^{\ast}\psi_{\mathbf{z}}%
\psi_{\mathbf{y}}\right\rangle , \label{U}%
\end{equation}
where $\left\langle ...\right\rangle =\mathrm{tr}\left(  ...\rho\right)  $.
The effective interaction $v^{\mathrm{eff}}(\mathbf{y,z})$ is \emph{defined}
by the relation
\begin{equation}
\int d\mathbf{y}d\mathbf{z}v(|\mathbf{y}-\mathbf{z}|)\left\langle
\psi_{\mathbf{y}}^{\ast}\psi_{\mathbf{z}}^{\ast}\psi_{\mathbf{z}}%
\psi_{\mathbf{y}}\right\rangle =\int d\mathbf{y}d\mathbf{z}v^{\mathrm{eff}%
}(\mathbf{y,z})\left(  \rho_{\mathbf{y}}\rho_{\mathbf{z}}-\rho_{\mathbf{yz}%
}\rho_{\mathbf{zy}}\right)  . \label{eff}%
\end{equation}
Then on using (\ref{rxy}), Eq. (\ref{U}) becomes
\begin{equation}
U=\frac{1}{2m}\sum_{\lambda}\int d\mathbf{y}f_{\lambda}\left[  \nabla
\varphi_{\lambda}\left(  \mathbf{y}\right)  \right]  ^{2}+\frac{1}{2}\int
d\mathbf{y}d\mathbf{z}v^{\mathrm{eff}}(\mathbf{y,z})\left(  \rho_{\mathbf{y}%
}\rho_{\mathbf{z}}-\rho_{\mathbf{yz}}\rho_{\mathbf{zy}}\right)  . \label{Ua}%
\end{equation}

Assuming that $v^{\mathrm{eff}}(\mathbf{x,y})$ does not depend on
$\rho_{\mathbf{R}}$, where $\mathbf{R=}\frac{1}{2}(\mathbf{x+y)}$, we find
from (\ref{d(U-)}), (\ref{Ua}) the infinite system of equations%
\begin{equation}
-\frac{1}{2m}\nabla^{2}\varphi_{\nu}\left(  \mathbf{x}\right)  +\left(
V^{\mathrm{HF}}+\Lambda Y_{\mathbf{x}}\right)  \varphi_{\nu}\left(
\mathbf{x}\right)  =E_{\nu}\varphi_{\nu}\left(  \mathbf{x}\right)  ,
\label{Sch}%
\end{equation}
where $E_{\nu}=e_{\nu}+\mu$ and
\begin{equation}
V^{\mathrm{HF}}\varphi_{\nu}\left(  \mathbf{x}\right)  =\int d\mathbf{y}%
v^{\mathrm{eff}}(\mathbf{x,y})\left[  \rho_{\mathbf{y}}\varphi_{\nu}\left(
\mathbf{x}\right)  -\rho_{\mathbf{xy}}\varphi_{\nu}\left(  \mathbf{y}\right)
\right]  , \label{VHF}%
\end{equation}
which constitutes, in combination with equations (\ref{rxy}), (\ref{ryHF}) for
$\rho_{\mathbf{y}}$ and $\rho_{\mathbf{xy}}$, the formal basis of the
temperature-dependent Hartree-Fock method.

In order to account for the fact that $v^{\mathrm{eff}}(\mathbf{x,y})$ does
depend on $\rho_{\mathbf{R}}$, it is necessary to specify the exact form for
such a dependence. For instance, in the case, when $v^{\mathrm{eff}%
}(\mathbf{x,y})$ is the Skyrme force $v_{\mathrm{Sky}}(\mathbf{x,y})$, Eq.
(\ref{Ua}) takes the form $U_{\mathrm{Sky}}=\int d\mathbf{x}H_{\mathrm{Sky}%
}\left(  \mathbf{x}\right)  $, in which $H_{\mathrm{Sky}}\left(
\mathbf{x}\right)  $ is a certain algebraic function of the number density
$\rho_{\mathbf{x}}$, the kinetic energy density $2m\tau_{\mathbf{x}}$, the
spin-orbit density $\mathbf{J}_{\mathbf{x}}$, and their derivatives
\cite{BGH}. The equations for $\varphi_{\nu}\left(  \mathbf{x}\right)  $ will
then be the same as (\ref{Sch}), while (\ref{VHF}) takes the form
\begin{equation}
V_{\mathrm{Sky}}^{\mathrm{HF}}\varphi_{\lambda}\left(  \mathbf{x}\right)
=\nabla f_{1}\nabla\varphi_{\lambda}\left(  \mathbf{x}\right)  +f_{2}%
\varphi_{\lambda}\left(  \mathbf{x}\right)  , \label{VHFSky}%
\end{equation}
where $f_{1}$ depends on $\rho_{\mathbf{x}}$ only, whereas $f_{2}$ is an
algebraic function of $\rho_{\mathbf{x}}$, $\tau_{\mathbf{x}}$, $\mathbf{J}%
_{\mathbf{x}}$, and their derivatives.

The solutions $\varphi_{\nu}\left(  \mathbf{x}\right)  $ of Eqs. (\ref{Sch}),
(\ref{VHF}) with $\rho_{\mathbf{y}}$, $\rho_{\mathbf{xy}}$ defined in
(\ref{ryHF}), (\ref{rxy}) depend on $\beta$, $\mu$, $\Lambda$: $\rho
_{\mathbf{x}}\equiv\rho\left(  \mathbf{x};\beta,\mu,\Lambda\right)  $,
$\rho_{\mathbf{xy}}\equiv\rho\left(  \mathbf{x,y};\beta,\mu,\Lambda\right)  $.
Given $\rho_{\mathbf{x}}\equiv\rho\left(  \mathbf{x};\beta,\mu,\Lambda\right)
$ the chemical potential $\mu_{g}$ and deformation force $f=\Lambda_{g}$ as
functions of $\beta$, $A$, $q$, are determined from Eq. (\ref{mL}). Expansions
for $\tau_{\mathbf{x}}$ and $\mathbf{J}_{\mathbf{x}}$ in powers of $\hbar$,
involving $\rho_{\mathbf{x}}$ and its spatial derivatives, which are
converging at arbitrary $\mathbf{x}$, may be obtained at finite temperatures
only \cite{BGH,BB}. Using few lowest in $\hbar$ terms of those expansions in
$H_{\mathrm{Sky}}\left(  \mathbf{x}\right)  $, one gets the so-called extended
Thomas-Fermi (ETF) internal energy density $H_{\mathrm{Sky}}^{\mathrm{ETF}%
}\left(  \mathbf{x}\right)  $, which is a function of $\rho_{\mathbf{x}}$ and
its spatial derivatives: $H_{\mathrm{Sky}}^{\mathrm{ETF}}\left(
\mathbf{x}\right)  =H_{\mathrm{Sky}}^{\mathrm{ETF}}\left(  \rho_{\mathbf{x}%
}\mathbf{,}\nabla\rho_{\mathbf{x}},..\right)  $. The internal energy in this
case becomes a functional of $\rho_{\mathbf{x}}$ only, which allows to use the
extremum property of $\Omega\left[  \rho\right]  $ to write the variational
equation for $\rho_{\mathbf{x}}$%
\begin{equation}
\frac{\delta}{\delta\rho_{\mathbf{x}}}\left(  \int d\mathbf{y}H_{\mathrm{Sky}%
}^{\mathrm{ETF}}\left(  \rho_{\mathbf{y}}\mathbf{,}\nabla\rho_{\mathbf{y}%
},..\right)  +\Lambda\int d\mathbf{y}Y_{\mathbf{y}}\rho_{\mathbf{y}}-\mu\int
d\mathbf{y}\rho_{\mathbf{y}}\right)  =0. \label{B}%
\end{equation}
This equation has the same form as the Strutinsky-Tyapin equation (\ref{ve})
commented in Introduction. According to \cite{BGH}, Eq. (\ref{B}) enables one
to calculate $\rho_{\mathbf{x}}$ and $U$ two orders of magnitude faster than
the HF formalism. Moreover, the potential $V_{\mathrm{Sky}}^{\mathrm{HF}%
}=\nabla f_{1}\nabla+f_{2}$ obtained with ETF values of $\rho_{\mathbf{x}}$,
$\tau_{\mathbf{x}}$, $\mathbf{J}_{\mathbf{x}}$, being used in the HF
equations, provides a rather inexpensive and yet accurate shortcut to fully
quantum HF calculations.

For $\beta^{-1}\rightarrow+0$, the $f_{\lambda}$ goes to the step function and
Eqs. (\ref{Sch}), (\ref{VHF}) convert into the system of $A$ usual HF
equations, providing a formal basis of self-consistent mean field models.

As stressed in \cite{FQVVK} and \cite{ST,Tyapin71}, in contrast to the liquid
drop model, the HF formalism and its semiclassical version, Eq. (\ref{B}), are
capable to account for though small but finite compressibility of the nucleus.
In order to illustrate this, consider the density distributions $\rho
_{\mathbf{x}}$ obtained in ETF calculations \cite{ASCV,ACVN} performed for
$^{160}$Yb with a realistic Skyrme interaction, namely SkM$^{\ast}$
\cite{BGH}. The $Y_{\mathbf{x}}$ function was taken in the form $Y_{\mathbf{x}%
}=2z^{2}-r^{2}$, with $r=\sqrt{x_{1}^{2}+x_{2}^{2}}$,\ \ $z=x_{3}$, where
$x_{i}$ are the Cartesian coordinates of $\mathbf{x}$. The collective
coordinate $q$ for such a choice of $Y_{\mathbf{x}}$ has the meaning of the
quadrupole moment $q=Q_{2}$. Since $Y_{\mathbf{x}}$ is invariant under
rotations of $\mathbf{x}$ about the $x_{3}=z$ axis, the densities
$\rho_{\mathbf{x}}$ are functions of $z$, $r$: $\rho_{\mathbf{x}}=\rho(z,r)$.
The values of $\rho(z,r)/\rho_{0}$, with $\rho_{0}=A/(\frac{4}{3}\pi R_{0}%
^{3})$, $R_{0}=1.18\,A^{1/3}$ fm, along the fission path of $^{160}$Yb are
depicted in Fig. 1. The profile function $\rho_{c}(z)$ is defined as such
contour on the $\,z$, $r$ plot of $\rho(z,r)$, where $\rho(z,r)/\rho_{0}=0.5$.
In Fig. 2 one can see those $\rho_{c}(z)$ for the sequence of $\rho(z,r)$,
presented in Fig. 1. Let $V=V\left(  Q_{2}\right)  $ be the volume of a body
of constant density $\rho_{0}$ bound by the surface of revolution of $\rho
_{c}(z)$ about axis $z$. From Fig. 3 one can see that this volume slightly
decreases with elongation, which means that on its way to scission the nucleus
is getting denser. Using the empirical formula (see Eq. (2.70) in \cite{BM1})%
\[
R\approx(1.12A^{1/3}-0.86A^{-1/3})\text{ fm}%
\]
for nuclear radii $R$, we find that the ratio of the total volume
$2V_{\mathrm{ff}}$ of two spherical fission fragments to the volume
$V_{\mathrm{cn}}$ of the spherical compound nucleus $^{160}$Yb is
$2V_{\mathrm{ff}}/V_{\mathrm{cn}}=0.954$. This number is surprisingly close to
the value $V(240\mathrm{b})/V(0)\approx0.95$ from Fig. 3.

\subsection{Inertia}

By using (\ref{ryHF}), Eq. (\ref{Mq}) for inertia $M$ may be rewritten as
\begin{equation}
M=m\sum\limits_{\lambda}f_{\lambda}\int d\mathbf{x}\varphi_{\lambda}%
^{2}\left(  \mathbf{x}\right)  \widetilde{u}_{\mathbf{x}}^{2}=m\sum
\limits_{\lambda}f_{\lambda}\left\langle \lambda\right\vert \widetilde{u}%
^{2}\left\vert \lambda\right\rangle , \label{M0}%
\end{equation}
where $\widetilde{u}^{2}$ is the one-body operator with the matrix elements
$\left\langle \mathbf{x}\right\vert \widetilde{u}^{2}\left\vert \mathbf{y}%
\right\rangle =\widetilde{u}_{\mathbf{x}}^{2}\delta(\mathbf{x-y})$. Since
$\widetilde{\mathbf{u}}_{\mathbf{x}}=\nabla\widetilde{\phi}_{\mathbf{x}}$, the
inertia is expected to be expressible in terms of $\widetilde{\phi}%
_{\lambda\nu}=\int d\mathbf{x}\varphi_{\lambda}\left(  \mathbf{x}\right)
\widetilde{\phi}_{\mathbf{x}}\varphi_{\nu}\left(  \mathbf{x}\right)  $. The
derivation of such an expression begins with using (\ref{fl}), (\ref{Sch}) to
represent (\ref{M0}) in the form%
\begin{equation}
M=m\sum\limits_{\lambda}\left\langle \lambda\right\vert \widetilde{u}%
^{2}\left(  1+e^{\beta(\hat{H}-\mu)}\right)  ^{-1}\left\vert \lambda
\right\rangle . \label{Msh1}%
\end{equation}
Here, $\hat{H}$ is the single-particle Hamiltonian
\[
\hat{H}=\frac{1}{2m}\hat{\mathbf{p}}\hat{\mathbf{p}}+V_{\mathbf{x}},
\]
with $\hat{\mathbf{p}}=-i\nabla$ and
\begin{equation}
V_{\mathbf{x}}=V_{\mathrm{Sky}}^{\mathrm{HF}}+\Lambda Y_{\mathbf{x}},
\label{Vx}%
\end{equation}
in accordance with (\ref{Sch}), (\ref{VHFSky}). In the subsequent discussion
the $\nabla f_{1}\nabla$ term in $V_{\mathrm{Sky}}^{\mathrm{HF}}$ is for
simplicity omitted. Now, using the identity
\begin{equation}
\left[  \hat{\mathbf{p}},\widetilde{\phi}_{\mathbf{x}}\right]  =-i\left(
\nabla\widetilde{\phi}_{\mathbf{x}}\right)  , \label{id}%
\end{equation}
we obtain
\begin{equation}
\lbrack\hat{H},\widetilde{\phi}_{\mathbf{x}}]=\frac{-i}{2m}\left[  \left(
\nabla\widetilde{\phi}_{\mathbf{x}}\right)  \hat{\mathbf{p}}+\hat{\mathbf{p}%
}\left(  \nabla\widetilde{\phi}_{\mathbf{x}}\right)  \right]  \label{Fiwd}%
\end{equation}
and%
\begin{equation}
\lbrack\lbrack\hat{H},\widetilde{\phi}_{\mathbf{x}}],\widetilde{\phi
}_{\mathbf{x}}]=-\frac{1}{m}\left(  \nabla\widetilde{\phi}_{\mathbf{x}%
}\right)  ^{2}. \label{Hpp}%
\end{equation}
Next, replacing $\nabla\widetilde{\phi}_{\mathbf{x}}$ in (\ref{Hpp}) with
$\widetilde{\mathbf{u}}_{\mathbf{x}}$ in accordance with (\ref{pot}), we have
\begin{equation}
\widetilde{u}_{\mathbf{x}}^{2}=-m[[\hat{H},\widetilde{\phi}_{\mathbf{x}%
}],\widetilde{\phi}_{\mathbf{x}}]. \label{Hff}%
\end{equation}
Substituting (\ref{Hff}) into (\ref{Msh1}) and accounting for (\ref{Sch}), we
find the desired expression for $M$:
\begin{equation}
M=-m^{2}\sum\limits_{\lambda\nu}\left\vert \widetilde{\phi}_{\lambda\nu
}\right\vert ^{2}\left(  f_{\lambda}-f_{\nu}\right)  E_{\lambda\nu},
\label{Mqfi}%
\end{equation}
where $E_{\lambda\nu}\equiv E_{\lambda}-E_{\nu}$.

For calculating $M$ with the aid of (\ref{Mqfi}) one should solve Eq.
(\ref{icq}) for $\widetilde{\phi}_{\mathbf{x}}$, which in the nuclei of
complex shape is quite tedious. The well-known simplification consists in
using the Werner-Wheeler approximation for $\widetilde{\mathbf{u}}%
_{\mathbf{x}}$. This approximation does not satisfy the condition
$\widetilde{\mathbf{u}}_{\mathbf{x}}=\nabla\widetilde{\phi}_{\mathbf{x}}$,
hence there is no Werner-Wheeler version of the phase function $\widetilde
{\phi}_{\mathbf{x}}$, to be used in (\ref{Mqfi}). To circumvent this obstacle
we express $M$ via the matrix elements of the one-body operator
\begin{equation}
\hat{F}=i\left[  \hat{H},\frac{1}{2}(\widetilde{\mathbf{u}}_{\mathbf{x}}%
\hat{\mathbf{p}}+\hat{\mathbf{p}}\widetilde{\mathbf{u}}_{\mathbf{x}}%
\mathbf{)}\right]  . \label{Fwspq}%
\end{equation}
Replacing $\nabla\widetilde{\phi}_{\mathbf{x}}$ in the right-hand side of
(\ref{Fiwd}) with $\widetilde{\mathbf{u}}_{\mathbf{x}}$, we find the formula
\[
\frac{i}{2}(\widetilde{\mathbf{u}}_{\mathbf{x}}\hat{\mathbf{p}}+\hat
{\mathbf{p}}\widetilde{\mathbf{u}}_{\mathbf{x}}\mathbf{)=-}m[\hat
{H},\widetilde{\phi}_{\mathbf{x}}],
\]
which allows to put Eq. (\ref{Fwspq}) for $\hat{F}$ into the form
\[
\hat{F}=-m\left[  \hat{H},[\hat{H},\widetilde{\phi}_{\mathbf{x}}]\right]  ,
\]
from which, using (\ref{Sch}), we find
\begin{equation}
F_{\lambda\nu}=-mE_{\lambda\nu}^{2}\widetilde{\phi}_{\lambda\nu}. \label{FFi}%
\end{equation}
Inserting (\ref{FFi}) into Eq. (\ref{Mqfi}), we arrive at the required
expression for $M$ in terms of $F_{\lambda\nu}$:
\begin{equation}
M=-\sum\limits_{\lambda\nu}\left\vert F_{\lambda\nu}\right\vert ^{2}\left(
f_{\lambda}-f_{\nu}\right)  /E_{\lambda\nu}^{3}. \label{Mc}%
\end{equation}

In order to use (\ref{Mc}) in practical calculations, Eq. (\ref{Fwspq}) for
$\hat{F}$ is rewritten in the form
\begin{equation}
\hat{F}=\hat{F}^{(V)}+\hat{F}^{(u)}, \label{F}%
\end{equation}
with
\begin{equation}
\hat{F}^{(V)}=-\widetilde{\mathbf{u}}_{\mathbf{x}}\nabla V_{\mathbf{\mathbf{x}%
}} \label{hFV}%
\end{equation}
and%
\begin{equation}
\hat{F}^{(u)}=\frac{1}{4m}\left(  \hat{p}_{i}\hat{p}_{k}\widetilde{u}%
^{ik}+\hat{p}_{i}\widetilde{u}^{ik}\hat{p}_{k}+\hat{p}_{k}\widetilde{u}%
^{ik}\hat{p}_{i}+\widetilde{u}^{ik}\hat{p}_{i}\hat{p}_{k}\right)  ,
\label{hFu}%
\end{equation}
where $\widetilde{u}^{ik}\equiv\partial\left(  \widetilde{u}_{\mathbf{x}%
}\right)  _{i}/\partial x_{k}$. The Latin indices represent the Cartesian
components of the corresponding vector. The repeated indices are summed over
from 1 to 3.

Following \cite{KR}, consider a simplistic situation, when $\mathbf{u}%
_{\mathbf{x}}$ and $V_{\mathbf{\mathbf{x}}}$ are given by:

1) $\mathbf{u}_{\mathbf{x}}$ $=0$ if $\mathbf{\mathbf{x}}$ is off the nuclear
surface and $\mathbf{u}_{\mathbf{a}}=\dot{q}\mathbf{n}_{\mathbf{a}}$ if
$\mathbf{\mathbf{x=a}}$, where $\mathbf{a}$ is an arbitrary point of the
nuclear surface and $\mathbf{n}_{\mathbf{a}}$ is a unit vector normal to the
surface at point $\mathbf{a}$;

2) $V_{\mathbf{\mathbf{x}}}=$ $-V_{0}$ safely inside the nucleus,
$V_{\mathbf{\mathbf{x}}}=$ 0 safely outside of it, and $V_{\mathbf{\mathbf{x}%
}}$ is a step-like function of $\zeta\mathbf{n}_{\mathbf{a}}$ with vanishing
width for $\mathbf{\mathbf{x}}$ close to the surface, where $\zeta=\left(
\mathbf{\mathbf{x-}a}\right)  \mathbf{n}_{\mathbf{a}}$ and $\mathbf{a}$ is
that point on the nuclear surface, which is the most close to
$\mathbf{\mathbf{x}}$.

It can be shown, using Eqs. (\ref{uw}), (\ref{F}), (\ref{hFV}), (\ref{hFu}),
that $\hat{F}$ goes to $\partial_{q}V_{\mathbf{\mathbf{x}}}$ in this case, so
that $M$ given in (\ref{Mc}) goes to
\begin{equation}
M^{\mathrm{cr}}=-\sum\limits_{\lambda\nu}\left\vert (\partial_{q}%
V)_{\lambda\nu}\right\vert ^{2}\left(  f_{\lambda}-f_{\nu}\right)
/E_{\lambda\nu}^{3}.\label{Dq}%
\end{equation}
The index $\mathrm{cr}$ is designed to stress that Eq. (\ref{Dq}), according
to \cite{YHS}, follows from the cranking model formula for inertia of a Fermi
gas \cite{H}, if any residual interaction is neglected. It can be shown
analytically and numerically calculations \cite{YHS} that $M^{\mathrm{cr}}$ as
a function of $q$ is divergent near the crossing points of the single-particle
energy levels, where $E_{\lambda\nu}=E_{\lambda}-E_{\nu}=0$. This signals that
the simplistic situation considered in \cite{KR} is physically meaningless.
Our derivation of (\ref{Dq}) shows that $M^{\mathrm{cr}}$ could have been
applicable for such a hypothetical nucleus, in which the mass flow induced by
the distortion of the nuclear surface is getting immediately damped. This
implies, of course, a very short mean free path, which is incompatible with
the Fermi gas model of the nucleus used to derive (\ref{Dq}).

The derivation of $M^{\mathrm{cr}}$ given in \cite{H,YHS} is based on the
linear response theory. The fact that this theory leads to wrong inertia means
that its application to nuclear collective motion is inappropriate. Really,
the LRT regards such motion as a sequence of the states of \emph{complete}
thermal equilibrium at fixed nuclear shape. Within this simplistic picture
there is no room for the \emph{mass} \emph{flow} in a hot time-evolving
nucleus, which precludes the possibility for LRT to get the realistic
estimates not only of collective inertia, but also of friction (see
\cite{Aleshin07} for details).

In passing we note that some formally erroneous simplifications, destroying
the self-consistency, need also to be made in deriving the cranking formula in
cold nuclei from the adiabatic time-dependent Hartree-Fock theory (see Section
9.3 of \cite{BV} for details). Though the cranking model expression for the
inertia in the dynamic equation for $q$ is wrong, the cranking model formulae
for the inertia of the rotational and translational motions are correct. This
is because the operators, which generate rotations and translations commute
with the total Hamiltonian, while the operator $P$ of collective moment
conjugated to $Q$ does not.

\subsection{Friction}

Let $a_{\lambda}^{\ast}$, $a_{\lambda}$ represent the creation and
annihilation operators of quasiparticles with energies $E_{\lambda}$ and wave
functions $\varphi_{\nu}\left(  \mathbf{x}\right)  $, as found from Eqs.
(\ref{Sch}), (\ref{VHFSky}). Substituting the expressions%
\begin{equation}
\psi{}_{\mathbf{x}}=\sum_{\nu}a_{\nu}\varphi_{\nu}\left(  \mathbf{x}\right)
,\text{ \ \ \ \ }\psi{}_{\mathbf{x}}^{\ast}=\sum_{\lambda}a_{\lambda}^{\ast
}\varphi_{\lambda}\left(  \mathbf{x}\right)  \label{psia}%
\end{equation}
into (\ref{gf}) and using (\ref{Cff}), we find
\begin{equation}
\gamma=im^{2}\sum_{\nu\lambda\nu_{1}\lambda_{1}}\widetilde{\phi}_{\lambda\nu
}\widetilde{\phi}_{\lambda_{1}\nu_{1}}\int_{-\infty}^{\infty}dte^{-\eta
|t|}t\frac{d^{4}}{dt^{4}}\left\langle a_{\lambda}^{\ast}\left(  t\right)
a_{\nu}\left(  t\right)  a_{\lambda_{1}}^{\ast}a_{\nu_{1}}\right\rangle _{g},
\label{g2}%
\end{equation}
where
\[
a_{\lambda}^{\ast}(t)=e^{it\mathcal{H}_{g}}a_{\lambda}^{\ast}e^{-it\mathcal{H}%
_{g}},\ \ \ \ \ \ \ \ a_{\nu}(t)=e^{it\mathcal{H}_{g}}a_{\nu}e^{-it\mathcal{H}%
_{g}}.
\]
For practical calculations of $\left\langle a_{\lambda}^{\ast}\left(
t\right)  a_{\nu}\left(  t\right)  a_{\lambda_{1}}^{\ast}a_{\nu_{1}%
}\right\rangle $ we take $\rho_{g}$ in the form
\[
\rho_{g}=Z_{g}^{-1}\exp\left[  -\beta\sum_{\lambda}\left(  E_{\lambda}%
-\mu\right)  a_{\lambda}^{\ast}a_{\lambda}\right]  .
\]
The Hamiltonian $\mathcal{H}_{g}$, which enters the propagators in
$C_{\widetilde{\phi}\widetilde{\phi}}\left(  t\right)  $, is modelled as%
\[
\mathcal{H}_{g}=\mathcal{H}_{0}+V^{\mathrm{res}},
\]
where
\begin{equation}
\text{\ }\mathcal{H}_{0}=\sum_{\lambda}E_{\lambda}a_{\lambda}^{\ast}%
a_{\lambda}=-\frac{1}{2m}\int d\mathbf{x}\psi{}_{\mathbf{x}}^{\ast}\nabla
^{2}\psi{}_{\mathbf{x}}+\int d\mathbf{x}\psi{}_{\mathbf{x}}^{\ast
}V_{\mathbf{x}}\psi{}_{\mathbf{x}} \label{H0}%
\end{equation}
is the mean-field Hamiltonian, while $V^{\mathrm{res}}$ is the residual
interaction. The second equality in (\ref{H0}) is obtained by using
(\ref{Sch}), (\ref{psia}). The $V^{\mathrm{res}}$ is given by
\[
V^{\mathrm{res}}=\frac{1}{2}\int d\mathbf{y}d\mathbf{z}v^{\mathrm{eff}%
}(\mathbf{y,z})\psi_{\mathbf{y}}^{\ast}\psi_{\mathbf{z}}^{\ast}\psi
_{\mathbf{z}}\psi_{\mathbf{y}}-\sum_{\lambda\nu}V_{\lambda\nu}^{\mathrm{HF}%
}a_{\lambda}^{\ast}a_{\nu},
\]
where
\[
V_{\lambda\nu}^{\mathrm{HF}}=\int d\mathbf{x}d\mathbf{y}\varphi_{\lambda
}\left(  \mathbf{x}\right)  v^{\mathrm{eff}}(\mathbf{x,y})\left[
\rho_{\mathbf{y}}\varphi_{\nu}\left(  \mathbf{x}\right)  -\rho_{\mathbf{xy}%
}\varphi_{\nu}\left(  \mathbf{y}\right)  \right]  .
\]
Using (\ref{ryHF}), (\ref{rxy}), we obtain%
\[
V_{\lambda\nu}^{\mathrm{HF}}=\sum_{\alpha}f_{\alpha}\int d\mathbf{x}%
d\mathbf{y}\varphi_{\lambda}\left(  \mathbf{x}\right)  \varphi_{\alpha}\left(
\mathbf{y}\right)  v^{\mathrm{eff}}(\mathbf{x,y})\left[  \varphi_{\nu}\left(
\mathbf{x}\right)  \varphi_{\alpha}\left(  \mathbf{y}\right)  -\varphi
_{\alpha}\left(  \mathbf{x}\right)  \varphi_{\nu}\left(  \mathbf{y}\right)
\right]  .
\]

Let $t_{0}$ represent the collision time: $t_{0}\sim$ $r_{0}/v_{F}$, where
$r_{0}$ is the range of the effective force and $v_{F}$ is the Fermi velocity.
Then, following \cite{SJH,Aleshin05}, we postulate that for $|t|>t_{0}$%
\begin{equation}
\left\langle a_{\lambda}^{\ast}\left(  t\right)  a_{\nu}\left(  t\right)
a_{\lambda_{1}}^{\ast}a_{\nu_{1}}\right\rangle =\left\langle a_{\lambda}%
^{\ast}\left(  t\right)  a_{\nu_{1}}\right\rangle \left\langle a_{\nu}\left(
t\right)  a_{\lambda_{1}}^{\ast}\right\rangle +\delta_{\nu\lambda}f_{\lambda
}\delta_{\nu_{1}\lambda_{1}}f_{\lambda_{1}}, \label{aaaa}%
\end{equation}
where%
\begin{equation}
\left\langle a_{\lambda}^{\ast}\left(  t\right)  a_{\nu_{1}}\right\rangle
=\delta_{\nu_{1}\lambda}f_{\lambda}e^{-\frac{1}{2}\Gamma_{\lambda}%
|t|}e^{iE_{\lambda}t},\text{ \ \ \ }\left\langle a_{\nu}\left(  t\right)
a_{\lambda_{1}}^{\ast}\right\rangle =\delta_{\nu\lambda_{1}}\bar{f}_{\nu
}e^{-\frac{1}{2}\Gamma_{\nu}|t|}e^{-iE_{\nu}t}, \label{g<>}%
\end{equation}
with \ $\bar{f}_{\nu}=1-f_{\nu}$. Within perturbative treatment of
$V^{\mathrm{res}}$, the collision rate $\Gamma_{\lambda}$ of quasiparticles
with nuclear media is related to the effective force as
\begin{equation}
\Gamma_{\mathbf{1}}=2\pi\sum_{\mathbf{234}}\left\langle \mathbf{12}\right\vert
v^{\mathrm{eff}}\left\vert \mathbf{34}\right\rangle \left\langle
\mathbf{43}\right\vert v^{\mathrm{eff}}\left\vert \mathbf{21}\right\rangle
_{A}\delta\left(  E_{\mathbf{1234}}\right)  \left(  f_{\mathbf{2}}\bar
{f}_{\mathbf{3}}\bar{f}_{\mathbf{4}}+\bar{f}_{\mathbf{2}}f_{\mathbf{3}%
}f_{\mathbf{4}}\right)  , \label{Ts}%
\end{equation}
where
\[
\left\vert \mathbf{21}\right\rangle _{A}=\left\vert \mathbf{21}\right\rangle
-\left\vert \mathbf{12}\right\rangle ,\ \ \ \ \ E_{\mathbf{1234}%
}=E_{\mathbf{1}}+E_{\mathbf{2}}-E_{\mathbf{3}}-E_{\mathbf{4}}.
\]
The bold digits for quasi-particle quantum numbers are used to stress that the
effective interaction $v^{\mathrm{eff}}$ in $\Gamma_{\mathbf{1}}$ is not
assumed to be diagonal over the spin-isospin quantum numbers $s$. If
$v^{\mathrm{eff}}$ does have this property,
\[
\left\langle s_{1}s_{2}\right\vert v^{\mathrm{eff}}(\mathbf{x},\mathbf{y}%
)\left\vert s_{3}s_{4}\right\rangle =\delta_{s_{1}s_{3}}\delta_{s_{2}s_{4}%
}v(\mathbf{x,y}),
\]
and then if $v(\mathbf{x,y})$ has the form
\[
v(\mathbf{x,y})=V_{0}(\mathbf{x})\delta(\mathbf{x}-\mathbf{y}),
\]
then $\Gamma_{\mathbf{1}}$ becomes
\[
\Gamma_{1}=6\pi\sum_{234}v_{1234}^{2}\delta\left(  E_{1234}\right)  \left(
f_{2}\bar{f}_{3}\bar{f}_{4}+\bar{f}_{2}f_{3}f_{4}\right)  ,
\]%
\[
v_{1234}=\int d\mathbf{x}\varphi_{1}\left(  \mathbf{x}\right)  \varphi
_{2}\left(  \mathbf{x}\right)  V_{0}(\mathbf{x})\varphi_{3}\left(
\mathbf{x}\right)  \varphi_{4}\left(  \mathbf{x}\right)  .
\]
The factor $6\pi$ instead of the usual $2\pi$ arises due to the fact that for
the $s$ independent zero-range interaction the summation over $s_{2}$, $s_{3}%
$, $s_{4}$ is reduced to finding the sum
\[
\sum_{s_{2}s_{3}s_{4}}\left\langle s_{1}s_{2}\right\vert \left\vert s_{3}%
s_{4}\right\rangle \left\langle s_{4}s_{3}\right\vert \left(  1-P^{\left(
s\right)  }\right)  \left\vert s_{2}s_{1}\right\rangle =3,
\]
where $P^{\left(  s\right)  }$ is the spin-isospin exchange operator.

Eqs. (\ref{aaaa}), (\ref{g<>}) imply that for $|t|>$ $t_{0}$ the trajectory of
any selected particle does not correlate with the trajectory of any selected
hole, unless their quantum numbers coincide. Both the particle and the hole,
however, have a chance to collide with nuclear media, which is reflected in
the damping factors in Eq. (\ref{g<>}) for their propagators. The simplest
derivation of Eqs. (\ref{g<>}), (\ref{Ts}), not involving a diagrammatic
expansion, but using instead Tserkovnikov's substitution for $\left\langle
a_{\lambda}^{\ast}\left(  t\right)  a_{\nu_{1}}\right\rangle _{g}$ and
$\left\langle a_{\nu}\left(  t\right)  a_{\lambda_{1}}^{\ast}\right\rangle
_{g}$, can be found in \cite{Aleshin03}. In \cite{Aleshin05} we use a simple
version of the effective force to obtain the semiclassical expression for
$\Gamma_{1}$ in finite spherical nuclei and compare it to the values of
$\Gamma_{1}$ in infinite matter.

The smallness of the collision time $t_{0}$ compared to the mean free path
time $\sim\Gamma^{-1}$, allows to use the asymptotic form (\ref{aaaa}),
(\ref{g<>}) of $\left\langle a_{\lambda}^{\ast}\left(  t\right)  a_{\nu
}\left(  t\right)  a_{\lambda_{1}}^{\ast}a_{\nu_{1}}\right\rangle _{g}$ for
all $t$ in (\ref{g2}), giving%
\[
\gamma=im^{2}\sum_{\nu\lambda}\widetilde{\phi}_{\lambda\nu}^{2}f_{\lambda}%
\bar{f}_{\nu}\int_{-\infty}^{\infty}dte^{-\eta|t|}t\frac{d^{4}}{dt^{4}}\left(
e^{-\Gamma_{\lambda\nu}|t|}e^{-iE_{\lambda\nu}t}\right)  ,
\]
where $E_{\nu\lambda}=E_{\nu}-E_{\lambda}$, $\Gamma_{\lambda\nu}=\frac{1}%
{2}\left(  \Gamma_{\lambda}+\Gamma_{\nu}\right)  $. Then, we use the formula%
\[
\lim_{\eta/\Gamma\rightarrow0}\int_{-\infty}^{\infty}dte^{-\eta|t|}%
t\frac{d^{4}}{dt^{4}}e^{-\Gamma|t|}e^{-i\omega t}=4i\omega\Gamma
\]
to obtain%
\[
\gamma=-4m^{2}\sum_{\nu\lambda}\widetilde{\phi}_{\lambda\nu}^{2}E_{\lambda\nu
}\Gamma_{\lambda\nu}f_{\lambda}\bar{f}_{\nu}.
\]
Noting that the $\widetilde{\phi}_{\lambda\nu}^{2}E_{\lambda\nu}%
\Gamma_{\lambda\nu}$ factor in the summand is antisymmetric under interchange
of $\lambda$ with $\nu$, we replace the remaining factor $f_{\lambda}\bar
{f}_{\nu}$ by its antisymmetric component%
\[
\left(  f_{\lambda}\bar{f}_{\nu}\right)  _{a}=\frac{1}{2}\left(  f_{\lambda
}\bar{f}_{\nu}-f_{\nu}\bar{f}_{\lambda}\right)  =\frac{1}{2}\left[
f_{\lambda}\left(  1-f_{\nu}\right)  -f_{\nu}\left(  1-f_{\lambda}\right)
\right]  =\frac{1}{2}\left(  f_{\lambda}-f_{\nu}\right)
\]
to obtain%
\begin{equation}
\gamma=-2m^{2}\Gamma\sum_{\nu\lambda}\widetilde{\phi}_{\lambda\nu}^{2}\left(
f_{\lambda}-f_{\nu}\right)  E_{\lambda\nu}, \label{gf1}%
\end{equation}
where $\Gamma$ is an 'effective' value of $\Gamma_{\nu\lambda}$, not depending
on the quantum numbers.

A seemingly different expression for $\gamma$ follows from (\ref{gc}), if we
replace there $\mathcal{F}$ with $\mathcal{F}_{0}\equiv\sum_{\lambda\nu
}F_{\lambda\nu}a_{\lambda}^{\ast}a_{\nu}$, where $F_{\lambda\nu}=\left\langle
\lambda\right\vert \hat{F}\left\vert \nu\right\rangle $, with $\hat{F}$
defined in (\ref{Fwspq}). The replacement of $\mathcal{F}$ with $\mathcal{F}%
_{0}$ is motivated by the assumption that $V^{\mathrm{res}}$ is small and it
leads to the quantity
\begin{equation}
\gamma_{0}=-\sum_{\nu\lambda}\left\vert F_{\lambda\nu}\right\vert ^{2}\left(
f_{\lambda}-f_{\nu}\right)  \frac{2E_{\lambda\nu}\Gamma}{\left(  E_{\lambda
\nu}^{2}+\Gamma^{2}\right)  ^{2}}. \label{eq36}%
\end{equation}
From Eqs. (\ref{F}), (\ref{hFV}), ( \ref{hFu}) for $\hat{F}$ it is seen that
Eq. (\ref{eq36}) coincides with the formula of Ref. \cite{Aleshin07}. Since
the assumption on the smallness of $V^{\mathrm{res}}$ implies that $\Gamma$ is
small too, we may omit $\Gamma$ in the denominator of (\ref{eq36}), which on
accounting for Eq. (\ref{FFi}) for $F_{\lambda\nu}$ leads to an expression for
$\gamma_{0}$, which is identical to (\ref{gf1}).

Comparing Eqs. (\ref{gf1}) and (\ref{Mqfi}) for $\gamma$ and $M$,
respectively, we find the formula%
\begin{equation}
\gamma=2\Gamma M, \label{gb}%
\end{equation}
which shows that the $q$ dependence of the reduced friction coefficient
$\gamma/M$ is completely determined by that of the collision rate $\Gamma$
defined in (\ref{Ts}). For practical use of this expression we rewrite it as%
\begin{equation}
\gamma=2\lambda^{-1}v_{\mathrm{F}}M, \label{gb1}%
\end{equation}
where $\lambda=v_{\mathrm{F}}/\Gamma$ is the mean free path of the nucleon.

In Figs. 4, 5 the $\gamma$ from (\ref{gb1}) are superimposed on the
semiclassical values $\gamma_{0}^{\mathrm{sc}}$ of $\gamma_{0}$ obtained from
(\ref{eq36}) in \cite{Aleshin07} by a conventional Monte-Carlo method. The
calculation is done for $^{208}$Pb and $^{272}$Ds at $\lambda=20$, $50$ fm
along the fission path starting from the spherical shape with the radius
$R=1.12A^{1/3}$ fm and ending at a necked-in configuration with the neck
radius of 0.3$R$. The $M$ given in (\ref{Mq}) was calculated for
$\rho_{\mathbf{x}}=$ 0.17 fm$^{-3}$. For $q$ we took the distance between the
mass centers of the two halves of the nucleus. The Werner-Wheeler expression
for the 'fluid velocity' $\widetilde{\mathbf{u}}_{\mathbf{x}}$ and the Cassini
oval parametrization for the profile function $\rho_{c}(z)$ of the nuclear
shape were used. One can see that $\gamma_{0}^{\mathrm{sc}}$ does reproduce
the general trends of $\gamma$ as a function of $A$, $\lambda$, and $q$,
although there are differences in detail. This may be related to the fact that
the summand for $\gamma_{0}^{\mathrm{sc}}$ is a sign-alternating function of
the initial position and velocity of the particle, which makes the direct
trajectory-integral simulations of $\gamma_{0}^{\mathrm{sc}}$ unreliable (e.g.
see \cite{MC}).

\section{Summary and some remarks}

The main objective of this study consists in working out a microscopic
description of collective motion in hot nuclei, allowing to express the
parameters of a phenomenological collective model \cite{BW,Kr,HW} in terms of
nucleonic quantities. The study starts with the introduction of the operator
$Q$ of collective coordinate and solving the two interwoven problems:
construction of the quasiequilibrium density matrix $\rho_{\mathrm{q}}$ and
establishing the explicit expressions for the operators $P$ and $\mathcal{M}$
of collective momentum conjugated to $Q$ and collective inertia, respectively.

The crucial steps are: 1) the substitution\ $\psi_{\mathbf{x}}\rightarrow
e^{-i\chi_{\mathbf{x}}}\psi_{\mathbf{x}}$,\ $\psi_{\mathbf{x}}^{\ast
}\rightarrow e^{i\chi_{\mathbf{x}}}\psi_{\mathbf{x}}^{\ast}$ for the nucleon
field operators made to get $\rho_{\mathrm{q}}$ from the canonical density
matrix $\rho_{g}$ subject to the condition $\mathrm{tr}\left(  Q\rho
_{g}\right)  =q$ and 2) the conversion of the adiabaticity hypothesis into
assumption that the deviations of the number density $\rho_{\mathbf{x},t}$ and
the momentum density $\mathbf{j}_{\mathbf{x},t}$, which are rapid functions of
time, from their slow components $\overline{\rho}_{\mathbf{x},t}$ and
$\overline{\mathbf{j}}_{\mathbf{x},t}$ defined in (\ref{rxt}) and (\ref{uxt}),
respectively, are smoothed out to a negligible level after performing the
$\mathbf{x}$ integrations invoked in the definitions of collective variables
$\mathrm{tr}\left(  Q_{t}\rho_{\mathrm{q}}\right)  $, $\mathrm{tr}\left(
P_{t}\rho_{\mathrm{q}}\right)  $, $\mathrm{tr}\left(  \mathcal{M}_{t}%
\rho_{\mathrm{q}}\right)  $ in terms of $\rho_{\mathbf{x},t}$ and
$\mathbf{j}_{\mathbf{x},t}$.

Equating two different expressions for the time derivative of $\mathrm{tr}%
\left(  K_{t}\rho_{\mathrm{q}}\right)  $ with $K_{t}=P_{t}-\dot{q}%
\mathcal{M}_{t}$ then led us to a relation between the variation $\delta
\dot{q}$ of collective velocity over the quasiequilibrium stage in question
and a time integral over this stage of the function $\mathrm{tr}\left(
i\left[  \mathcal{H},K_{t}\right]  \right)  $. After estimation of this
integral in the adiabatic approximation, the relation takes the same form as
the equation of motion of the phenomenological model, which allows us to
identify the many-body expressions for collective inertia $B$, deformation
force $f$, and friction coefficient $\gamma$.

Our microscopic expression $M$, (\ref{Mq}), for inertia $B$ has the form of
inertia of collective motion in an abstract fluid with the velocity
$\mathbf{\mathbf{\mathbf{\mathbf{\mathbf{u}}}}}_{\mathbf{x}}=\dot{q}%
\nabla\widetilde{\phi}_{\mathbf{x}}=m^{-1}\nabla\chi_{\mathbf{x}}$. It should
be stressed that $-m^{-1}\chi_{\mathbf{x}}$ has nothing to do with the
potential of the hydrodynamic velocity simply because this latter does not
make sense in the nuclear case, once the nucleon mean free path $\lambda$ is
comparable or greater than the nuclear sizes. To the best of our knowledge
this work is the first to derive the expression $f=$ $\Lambda_{g}$ for the
deformation force, where $\Lambda_{g}$ is the Lagrange multiplier used in the
expression for $\rho_{g}$ to constrain it to a given expectation value $q$ of
collective coordinate $Q$. Having shown that $\Lambda_{g}$ possesses the
property $\Lambda_{g}=-\partial_{q}F$, where $F$ is the free energy, we
establish another expression for the deformation force: $f=-\partial_{q}F$,
which though used in some analyses of collective motion, has never been
derived from first principles. Our microscopic expression for $\gamma$,
(\ref{gf}), relating this quantity with the time integral of a time
correlation function of $\widetilde{\phi}\equiv\int d\mathbf{x}\widetilde
{\phi}\mathbf{_{\mathbf{x}}}n_{\mathbf{x}}$ for the $\rho_{g}$ ensemble, is
also novel.

For practical calculations of $f$, $M$, and $\gamma$ the real nucleus is
replaced with a self-confined Fermi gas (SCFG) of quasiparticles, which,
unlike the standard gases confined by external forces, is confined by its own
forces. The SCFG model is based on the postulates that the one-body density
matrix and the entropy of strongly-interacting nucleons, represented by
$\rho_{g}$ have the same form as those of noninteracting fermions bound by an
external one-body potential and that the expectation of the bare interparticle
interaction for $\rho_{g}$ may be replaced with the expectation of the
\emph{effective }interaction for the two-body density matrix of
\emph{noninteracting} fermions. Use of the extreme property of Gibbs grand
potential constructed within the SCFG model for the $\rho_{g}$ ensemble allows
one to derive the temperature-dependent constrained Hartree-Fock equations
with the effective interaction. These equations become especially simple when
the effective interaction is the Skyrme force.

To find a SCFG expression for $\gamma$ from Eq. (\ref{gf}), we assume
independent decay of particle and hole states. Replacing for simplicity the
decay width $\Gamma_{\lambda}$ with the effective width $\Gamma$ not depending
on quantum numbers, we arrive at the formula $\gamma=2v_{\mathrm{F}}M/\lambda
$, where $v_{\mathrm{F}}$ is the Fermi velocity, $M$ is the collective
inertia, and $\lambda=v_{\mathrm{F}}/\Gamma$ the mean free path of the
nucleon. According to this formula the $A$, $q$, and temperature dependences
of the ratio $\gamma/M$ are completely determined by those of $v_{\mathrm{F}%
}/\lambda=$ $\Gamma$. Perhaps this result will foster further analytical and
computational studies of the nucleon mean free path $\lambda$ in spherical,
deformed, and superdeformed nuclei for realistic versions of the Skyrme force.

One of the principal results of this present paper and our previous work
\cite{Aleshin07} is that they stress that the linear response theory (LRT)
expressions for $B$ and $\gamma$
\cite{H,KHR,KR,K78,YHS,Aleshin99,Aleshin03,Aleshin05} are wrong. This is
because the LRT is not applicable in the nuclear case. Namely, one cannot
treat the force distorting the shape of a time evolving nucleus as an
\emph{external} force, because this force is depending on the \emph{intrinsic}
state of the nucleus. Moreover, the density matrix of the LRT implies that the
mass flow in the bulk of the time-evolving nucleus is absent.

To understand the nature of nuclear collective motion it was important to
realize a key role of the mass flow in this phenomenon and obtain an explicit
expression for this quantity in a slowly evolving nucleus: if the equilibrium
state of the nucleus is represented by the real-valued single-particle wave
functions $\varphi_{\lambda}\left(  \mathbf{x}\right)  $, for which the mass
flow
\[
\mathbf{j}_{\lambda}=-\left(  i/2\right)  (\varphi_{\lambda}^{\ast}%
\nabla\varphi_{\lambda}-\left(  \nabla\varphi_{\lambda}^{\ast}\right)
\varphi_{\lambda})=0,
\]
then in the quasiequilibrium state of the nucleus those wave functions acquire
the form $e^{i\chi_{\mathbf{x}}}\varphi_{\lambda}\left(  \mathbf{x}\right)  $,
to give rise to a nonvanishing mass flow
\[
\mathbf{j}_{\mathbf{x}}=\sum_{\lambda}f_{\lambda}\varphi_{\lambda}^{2}\left(
\mathbf{x}\right)  \nabla\chi_{\mathbf{x}},
\]
where $f_{\lambda}$ are the Fermi gas occupation numbers.

It is interesting to note that this way of generating the collective motion in
hot nuclei is similar to generating the coherent motion in the entrance
channel of the head-on nucleus-nucleus collision leading to the composite
system under study (the total angular momentum of the nucleus is 0 in our
model). Really, the ground states of two colliding nuclei can always be
represented by superpositions of Slater determinants of single-particle
standing waves. The relative motion with momentum $P_{\mathrm{r}}$ in the
head-on $A_{1}+A_{2}$ collision along the $x_{3}$ axis then gives rise to the
phase factor $e^{i\varphi_{\mathbf{x}}}$ attached to those standing waves
with
\[
\varphi_{\mathbf{x}}=-(P_{\mathrm{r}}/A_{1})x_{3}\text{ \ if}\ x_{3}%
<X_{3}\text{ \ \ and }\varphi_{\mathbf{x}}=(P_{\mathrm{r}}/A_{2})x_{3}\text{
\ \ if }x_{3}>X_{3},
\]
where $X_{3}$ is the touching point. This new vision of mechanism of
collective motion as manifestation of \emph{coherency of the phases} of
individual nucleons in the hot nucleus qualitatively differs from the N. Bohr
picture of collective motion in fission \cite{BW}, according to which the
collective motion arises due to \emph{coherency of the velocities} of
individual nucleons.

\section*{Acknowledgements}

The author is grateful to V.I. Abrosimov, M. Centelles, V.Yu. Denisov, S.N.
Fedotkin, V.M. Kolomietz, V.A. Plujko, and A.I. Sanzhur for useful discussions.

\section*{Appendix A}

In this appendix the operators $Q$, $\mathcal{M}$, $P$, $K$, $\mathcal{H}$,
$\dot{Q}$, $\dot{\mathcal{M}}$, $\dot{P}$, $\mathcal{H}_{\mathrm{c}}$, $V$,
$\dot{K}^{\mathrm{c}}$ are expressed in terms of the appropriate primed operators.

Using $n_{\mathbf{x}}=n_{\mathbf{x}}^{\prime}$ in (\ref{Qcoll}),
(\ref{Mcoll}), we obtain%
\begin{equation}
Q=Q^{\prime}, \label{QQ'}%
\end{equation}%
\begin{equation}
\mathcal{M}=\mathcal{M}^{\prime}. \label{MM'}%
\end{equation}
Eq. (\ref{Pcoll}) for $P$ on accounting for (\ref{pp'}), (\ref{uw}) takes the
form%
\begin{equation}
P=P^{\prime}+\dot{q}\mathcal{M}^{\prime}, \label{PP'}%
\end{equation}
where%
\begin{equation}
P^{\prime}\equiv\int d\mathbf{x\mathbf{\widetilde{\mathbf{u}}}%
_{\mathbf{\mathbf{x}}}\mathbf{\mathbf{\mathbf{\mathbf{p}}}}}_{\mathbf{x}%
}^{\prime},\text{ \ \ \ \ }\mathcal{M}^{\prime}\equiv m\int d\mathbf{x}%
\widetilde{u}_{\mathbf{\mathbf{x}}}^{2}n_{\mathbf{x}}^{\prime}. \label{P'M'}%
\end{equation}
Eq. (\ref{P2}) for $K$ after accounting for (\ref{PP'}), (\ref{MM'}) takes the
form%
\begin{equation}
K=P^{\prime}. \label{KK'}%
\end{equation}
Eq. (\ref{HH'}) for $\mathcal{H}$ on accounting for (\ref{uw}), (\ref{P'M'})
becomes
\begin{equation}
\mathcal{H}=\mathcal{H}^{\prime}+\text{\ }\dot{q}P^{\prime}%
\mathbf{\mathbf{\mathbf{\mathbf{+}}}}\frac{1}{2}\dot{q}^{2}\mathcal{M}%
^{\prime}. \label{HH'1}%
\end{equation}

Employing Eqs. (\ref{HH'1}), (\ref{QQ'}) in $\dot{Q}\equiv i\left[
\mathcal{H},Q\right]  $, we use $\left[  \mathcal{M}^{\prime},Q^{\prime
}\right]  =0$ to obtain \
\begin{equation}
\dot{Q}=i\left[  \mathcal{H}^{\prime}+\text{\ }\dot{q}P^{\prime}%
\mathbf{\mathbf{\mathbf{\mathbf{+}}}}\frac{1}{2}\dot{q}^{2}\mathcal{M}%
^{\prime},Q^{\prime}\right]  =i\left[  \mathcal{H}^{\prime}+\text{\ }\dot
{q}P^{\prime},Q^{\prime}\right]  =\dot{Q}^{\prime}-\dot{q}K_{1}^{\prime},
\label{QdQd'}%
\end{equation}
where $K_{1}^{\prime}=i\left[  Q^{\prime},P^{\prime}\right]  $. The equality
$\left[  \mathcal{M}^{\prime},Q^{\prime}\right]  =0$ follows from the relation
$\int d\mathbf{y}f_{\mathbf{y}}%
i[n_{\mathbf{\mathbf{\mathbf{\mathbf{\mathbf{\mathbf{y}}}}}}}^{\prime
},n_{\mathbf{x}}^{\prime}]=0$, valid for an arbitrary function $f_{\mathbf{y}%
}$. Employing Eqs. (\ref{HH'1}), (\ref{MM'}) in $\dot{\mathcal{M}}\equiv
i\left[  \mathcal{H},\mathcal{M}\right]  $, we find
\begin{equation}
\dot{\mathcal{M}}=i\left[  \mathcal{H}^{\prime}+\text{\ }\dot{q}P^{\prime
}\mathbf{\mathbf{\mathbf{\mathbf{+}}}}\frac{1}{2}\dot{q}^{2}\mathcal{M}%
^{\prime},\mathcal{M}^{\prime}\right]  =\dot{\mathcal{M}^{\prime}}-\dot
{q}K_{2}^{\prime}, \label{MdMd'}%
\end{equation}
where $K_{2}^{\prime}=i\left[  \mathcal{M}^{\prime},P^{\prime}\right]  $.
Employing Eqs. (\ref{HH'1}), (\ref{PP'}) in $\dot{P}\equiv i\left[
\mathcal{H},P\right]  $, we have \
\begin{equation}
\dot{P}=i\left[  \mathcal{H}^{\prime}+\text{\ }\dot{q}P^{\prime}%
\mathbf{\mathbf{\mathbf{\mathbf{+}}}}\frac{1}{2}\dot{q}^{2}\mathcal{M}%
^{\prime},P^{\prime}+\dot{q}\mathcal{M}^{\prime}\right]  =\dot{P}^{\prime
}+\dot{q}\dot{\mathcal{M}^{\prime}}-\frac{1}{2}\dot{q}^{2}K_{2}^{\prime}.
\label{PdPd'}%
\end{equation}

Eq. (\ref{Hcr}) for $\mathcal{H}_{\mathrm{c}}$ on accounting for Eqs.
(\ref{HH'1}), (\ref{PP'}), (\ref{QQ'}), (\ref{MM'}) takes the form
$\mathcal{H}_{\mathrm{c}}=\mathcal{H}^{\prime}+$\ $\Lambda_{g}Q^{\prime}$, or
using Eq. (\ref{Hg}) for $\mathcal{H}_{g}$,
\begin{equation}
\mathcal{H}_{\mathrm{c}}=\mathcal{H}_{g}^{\prime}. \label{Hcr4}%
\end{equation}
Eq. (\ref{V}) for $V$ on accounting for (\ref{QQ'}), (\ref{PP'}), (\ref{MM'})
becomes%
\begin{equation}
V=\text{\ }\Lambda_{g}Q^{\prime}-\text{\ }\dot{q}P^{\prime}-\frac{1}{2}\dot
{q}^{2}\mathcal{M}^{\prime}. \label{V'}%
\end{equation}
Eq. (\ref{Pndc}) for $\dot{K}^{\mathrm{c}}$ on accounting for (\ref{Hcr4}),
(\ref{KK'}) takes the form%
\begin{equation}
\dot{K}^{\mathrm{c}}=i\left[  \mathcal{H}_{g}^{\prime},P^{\prime}\right]
=\mathcal{F}^{\prime}, \label{Kdc}%
\end{equation}
where $\mathcal{F}^{\prime}$ is the primed partner of $\mathcal{F}$,
introduced in (\ref{Fc}).

\section*{Appendix B}

In this appendix we derive the integral equation (\ref{AP4}) for $\dot{K}%
_{t}=e^{it\mathcal{H}}\dot{K}e^{-it\mathcal{H}}$. The idea of the derivation
is taken from \cite{AP}. Define the operator
\[
J(t)\equiv e^{-it\mathcal{H}_{\mathrm{c}}}\dot{K}_{t}e^{it\mathcal{H}%
_{\mathrm{c}}},
\]
where $\dot{K}_{t}=e^{it\mathcal{H}}\dot{K}e^{-it\mathcal{H}}$. On using the
identities%
\begin{equation}
J(t)=J(0)+\left[  J(t)-J(0)\right]  =J(0)+\int_{0}^{t}ds\partial_{s}J(s),
\label{AP}%
\end{equation}%
\[
J(0)=\dot{K},
\]%
\[
\partial_{s}J(s)=\partial_{s}e^{-is\mathcal{H}_{\mathrm{c}}}e^{is\mathcal{H}%
}\dot{K}e^{-is\mathcal{H}}e^{is\mathcal{H}_{\mathrm{c}}}=e^{-is\mathcal{H}%
_{\mathrm{c}}}i\left[  \dot{K}_{s},V\right]  e^{is\mathcal{H}_{\mathrm{c}}},
\]
where $V=\mathcal{H}_{\mathrm{c}}-\mathcal{H}$, Eq. (\ref{AP}) becomes%
\[
e^{-it\mathcal{H}_{\mathrm{c}}}\dot{K}_{t}e^{it\mathcal{H}_{\mathrm{c}}}%
=\dot{K}+\int_{0}^{t}dse^{-is\mathcal{H}_{\mathrm{c}}}i\left[  \dot{K}%
_{s},V\right]  e^{is\mathcal{H}_{\mathrm{c}}},
\]
which leads in an evident way to the desired integral equation for $\dot
{K}_{t}$
\begin{equation}
\dot{K}_{t}=e^{it\mathcal{H}_{\mathrm{c}}}\dot{K}e^{-it\mathcal{H}%
_{\mathrm{c}}}+\int_{0}^{t}dse^{i(t-s)\mathcal{H}_{\mathrm{c}}}i\left[
\dot{K}_{s},V\right]  e^{-i(t-s)\mathcal{H}_{\mathrm{c}}}. \label{gen}%
\end{equation}

\section*{Appendix C}

Using the relation $\partial_{t}k_{n,t}=\left\langle {}\dot{K}_{n,t}%
\right\rangle _{\mathrm{q}}$ for $k_{n,t}\equiv\left\langle {}K_{n,t}%
\right\rangle _{\mathrm{q}}$ and Eqs. (\ref{iQP1}), (\ref{K2}), (\ref{rxt}),
we find
\begin{equation}
k_{n,t}=\int d\mathbf{x}w_{n,\mathbf{x}}\rho_{A,\beta}(\mathbf{x}%
,q_{t}),\text{ \ \ \ \ \ }w_{1,\mathbf{x}}=-\widetilde{\mathbf{u}}%
_{\mathbf{x}}\nabla Y\mathbf{_{\mathbf{x}},}\text{ \ \ \ \ \ }w_{2,\mathbf{x}%
}=-m\mathbf{\widetilde{\mathbf{u}}}_{\mathbf{\mathbf{x}}}\nabla\widetilde
{u}_{\mathbf{\mathbf{x}}}^{2}. \label{kit}%
\end{equation}
This equation says that $k_{n,t}$ are linear functions of $t$, which permits
us to write $\partial_{t}k_{n,t}=\delta k_{n}/\tau_{\mathrm{q}}$, with $\delta
k_{n}\equiv k_{n,\tau_{\mathrm{q}}}-k_{n,0}$. Thus, taking into account that
$\eta=\tau_{\mathrm{q}}^{-1}$, we obtain
\[
\int_{0}^{\infty}dte^{-\eta t}\left\langle {}\dot{K}_{n,t}\right\rangle
_{\mathrm{q}}=\int_{0}^{\infty}dte^{-t/\tau_{\mathrm{q}}}\partial_{t}%
k_{n,t}=\int_{0}^{\infty}dte^{-t/\tau_{\mathrm{q}}}\delta k_{n}/\tau
_{\mathrm{q}}=\delta k_{n}.
\]
From $\delta k_{n}=k_{n,\tau_{\mathrm{q}}}-k_{n,0}$ and (\ref{kit}) we find
the expressions
\begin{equation}
\delta k_{1}=-\delta q\cdot\int d\mathbf{x}\widetilde{\mathbf{u}}_{\mathbf{x}%
}\nabla Y\mathbf{_{\mathbf{x}}}\partial_{q}\rho_{\mathbf{x}},\text{
\ \ \ \ }\delta k_{2}=-\delta q\cdot m\left(  \int d\mathbf{x\widetilde
{\mathbf{u}}}_{\mathbf{\mathbf{x}}}\nabla\widetilde{u}_{\mathbf{\mathbf{x}}%
}^{2}\partial_{q}\rho_{\mathbf{x}}\right)  , \label{kit1}%
\end{equation}
where $\delta q=q_{\tau_{\mathrm{q}}}-q$. Since $\delta q=\dot{q}%
\tau_{\mathrm{q}}$, Eqs. (\ref{kit}), (\ref{kit1}) show that $\int_{0}%
^{\infty}dte^{-\eta t}\left\langle {}\dot{K}_{n,t}\right\rangle _{\mathrm{q}}$
are proportional to $\dot{q}$.

\section*{Appendix D}

Let us prove that $\left\langle i\left[  \mathcal{F}{},Q\right]  \right\rangle
_{g}=0$. Exploiting the cyclic property of the trace and definitions of
$\mathcal{F}{}$, $\mathcal{H}_{g}$, and $\dot{Q}$, we find that
\[
\left\langle i\left[  \mathcal{F}{},Q\right]  \right\rangle _{g}%
=\mathrm{tr}\left(  i\left[  \dot{Q},P{}\right]  \rho_{g}\right)  .
\]

From definitions of $Q$, $P$, $\dot{Q}$ and Eq. (\ref{nd}) it follows that
\[
\mathrm{tr}\left(  i\left[  \dot{Q},P{}\right]  \rho_{g}\right)
=\mathrm{tr}\left(  i\left[  \int d\mathbf{x}Y_{\mathbf{x}}i[\mathcal{H}%
,n_{\mathbf{x}}],\int d\mathbf{y}\widetilde{u}_{\mathbf{y}}^{\beta}%
{}p_{\mathbf{y}}^{\beta}\right]  \rho_{g}\right)  =
\]%
\[
=-m^{-1}\int d\mathbf{x}d\mathbf{y}Y_{\mathbf{x}}\nabla_{\mathbf{x}}^{\alpha
}\widetilde{u}_{\mathbf{y}}^{\beta}{}\mathrm{tr}\left(  i\left[
p_{\mathbf{x}}^{\alpha},p_{\mathbf{y}}^{\beta}\right]  \rho_{g}\right)  .
\]
From the relation%

\[
i\int d\mathbf{y}f_{\mathbf{\mathbf{\mathbf{\mathbf{\mathbf{\mathbf{y}}}}}}%
}[p_{\mathbf{\mathbf{\mathbf{\mathbf{\mathbf{\mathbf{y}}}}}}}^{\beta
},p_{\mathbf{x}}^{\alpha}]=-\frac{\partial(f_{\mathbf{x}}p_{\mathbf{x}%
}^{\alpha})}{\partial x_{\beta}}-\frac{\partial f_{\mathbf{x}}}{\partial
x_{\alpha}}p_{\mathbf{x}}^{\beta},
\]
valid for any c-number function
$f_{\mathbf{\mathbf{\mathbf{\mathbf{\mathbf{\mathbf{y}}}}}}}$, and Eq.
(\ref{pxg}), according to which $\mathrm{tr}\left(  \mathbf{p}_{\mathbf{x}%
}\rho_{g}\right)  =0$, it is seen that $\mathrm{tr}\left(  i\left[  \dot
{Q},P{}\right]  \rho_{g}\right)  =0$, hence, $\left\langle i\left[
\mathcal{F}{},Q\right]  \right\rangle _{g}=0$. Analogously one shows that
$\left\langle i\left[  \mathcal{F}{},\mathcal{M}\right]  \right\rangle _{g}=0$.

\begin{figure}[t]
\centerline{\includegraphics[width=0.95\textwidth,bb=28 0 424
429]{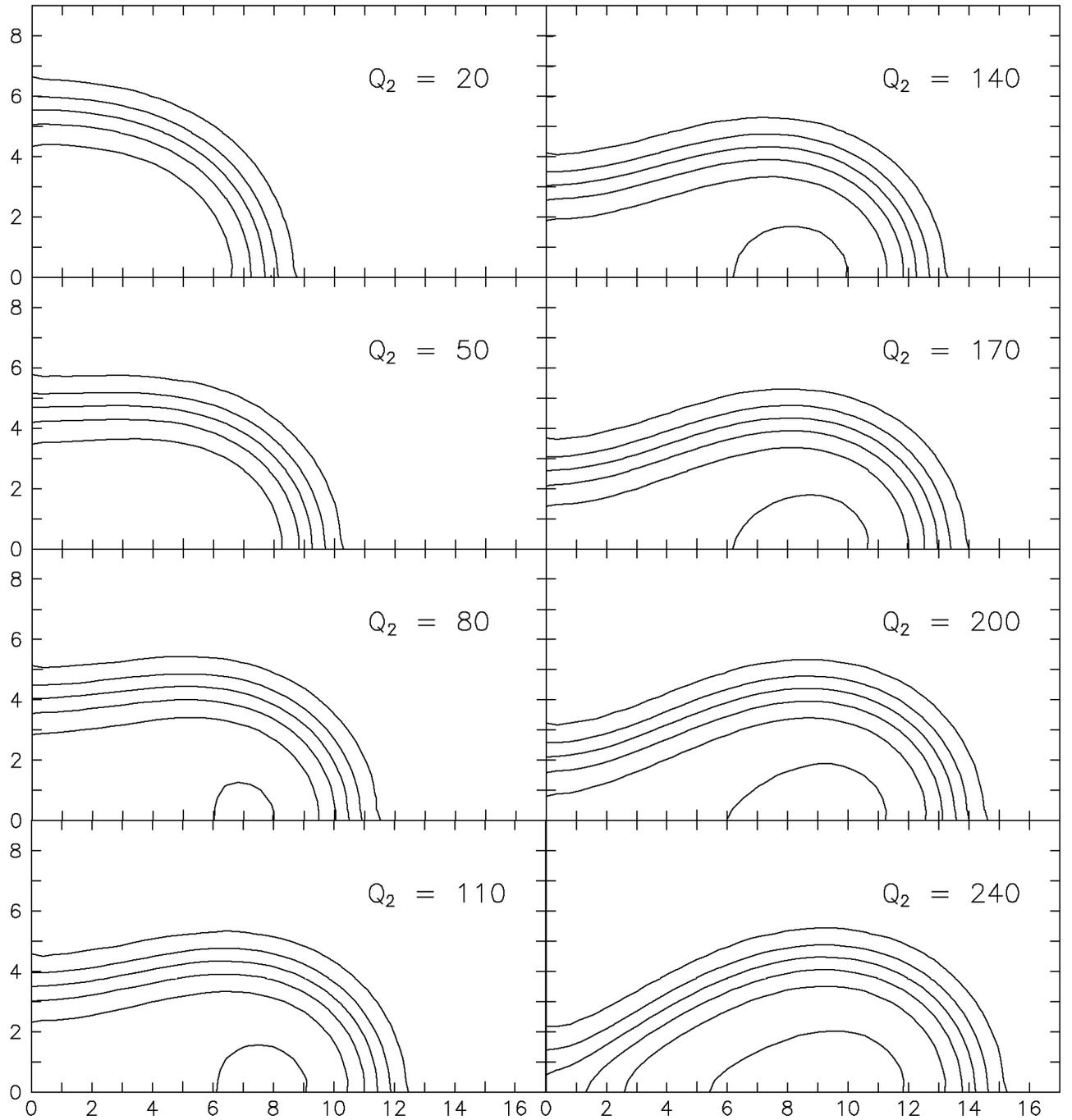}}\caption{Equidensity lines for $^{160}$Yb calculated by the
ETF method at the indicated values of the quadrupole moment (in barn). From
outside to inside, and in units of $\rho_{0}$, the lines represent contours of
constant density $\rho=$ 0.1, 0.3, 0.5, 0.7, 0.9 and 1.1. The horizontal axis
corresponds to the $z$ direction and the vertical axis to the $r$ direction
(units in fm). }%
\end{figure}

\begin{figure}[h]
\centerline{\includegraphics[width=0.75\textwidth,bb=0 15 276
229]{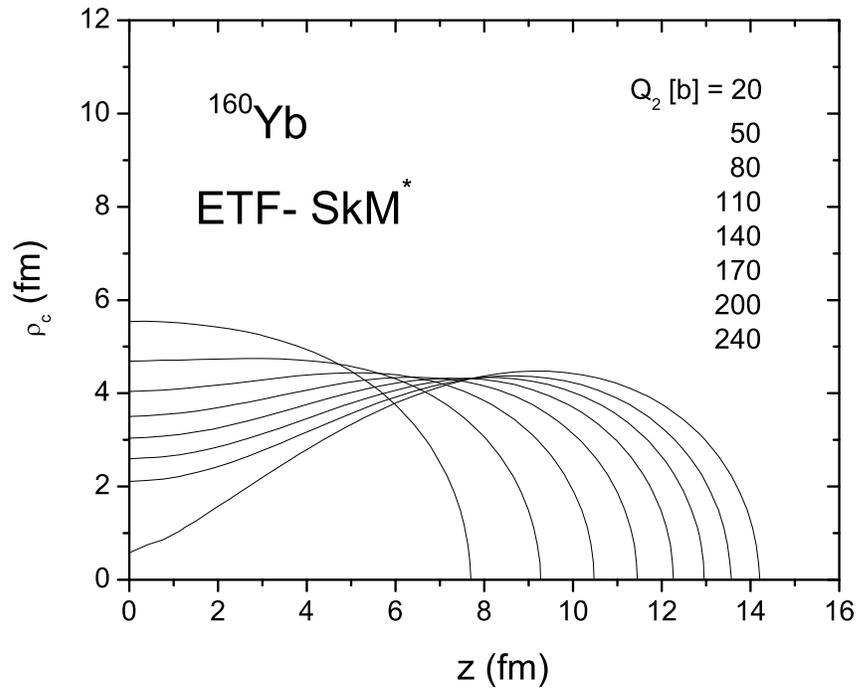}}\caption{Fission path for $^{160}$Yb as found from the ETF
density distributions.}%
\end{figure}

\begin{figure}[h]
\centerline{\includegraphics[width=0.75\textwidth,bb=0 15 276
233]{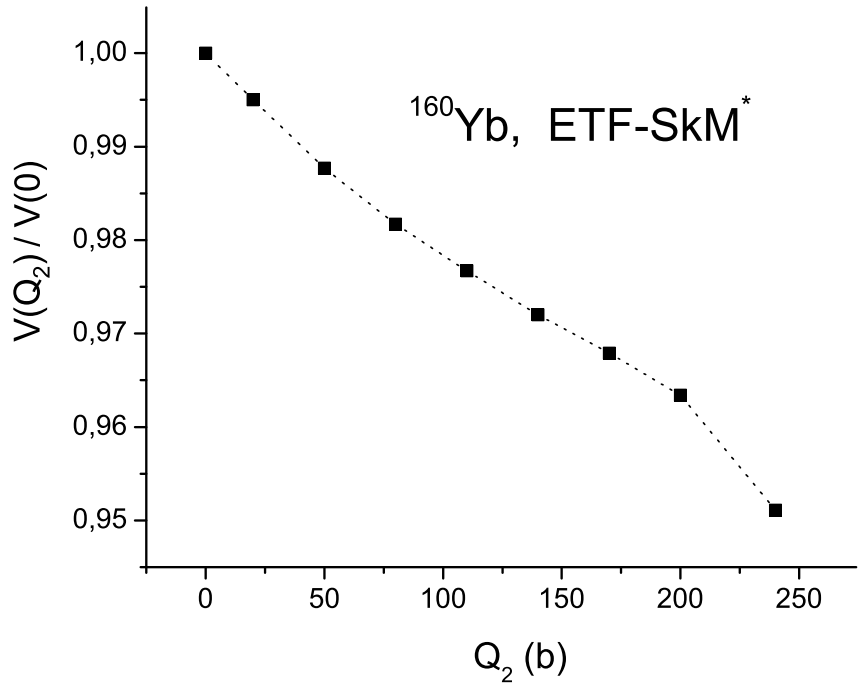}}\caption{The volume of $^{160}$Yb along the fission path in
units of its volume at $Q_{2}=0$ from the ETF - SkM$^{\ast}$ calculation. }%
\end{figure}

\begin{figure}[h]
\centerline{\includegraphics[width=0.75\textwidth,bb=0 15 276
233]{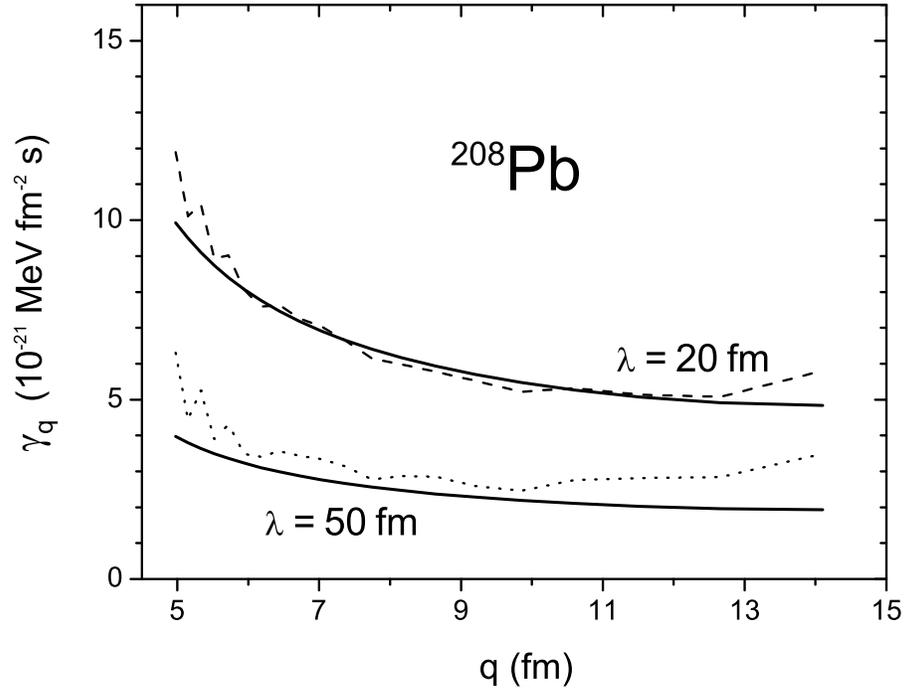}}\caption{Comparison between semiclassical friction
coefficients $\gamma_{0}^{\mathrm{sc}}$ (dotted and dashed) and $\gamma
=2v_{\mathrm{F}}M/\lambda$ (solid lines) along the fission path in $^{208}$Pb.
}%
\end{figure}

\begin{figure}[h]
\centerline{\includegraphics[width=0.75\textwidth,bb=0 15 276
233]{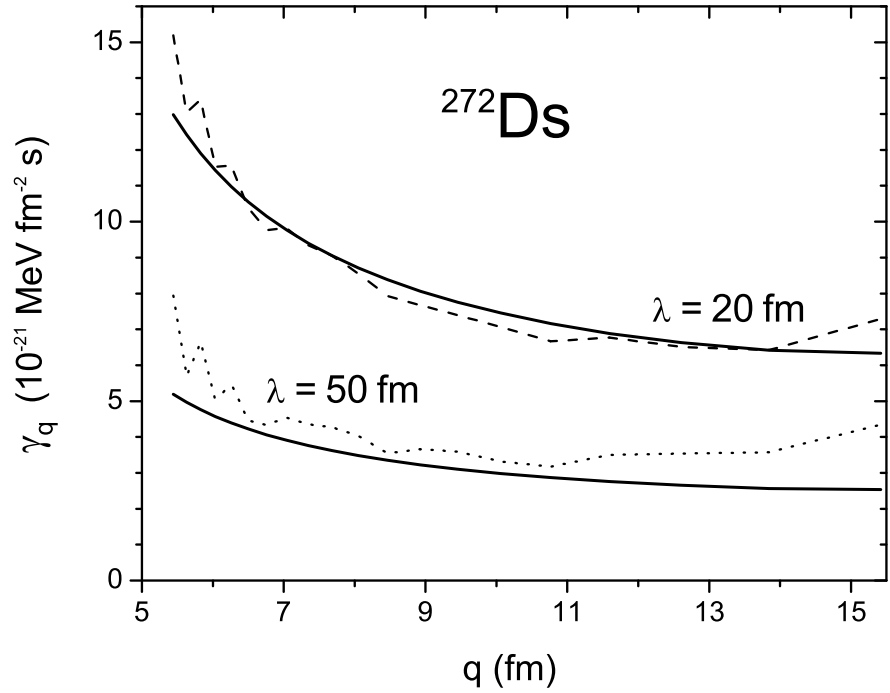}}\caption{Same as in the previous figure but for $^{272}$Ds.}%
\end{figure}

\end{document}